\documentclass[twocolumn, twocolappendix]{aastex631}
\usepackage{amsmath,amstext}
\usepackage{CJK}
\usepackage[T1]{fontenc}

\usepackage{newtxtext,newtxmath}
\usepackage[figure,figure*]{hypcap}
\usepackage{booktabs}
\usepackage{cleveref}
\usepackage{paralist}
\graphicspath{{fig/}}
\usepackage{ulem}

\usepackage{xcolor,colortbl}  % color for table

\usepackage[colorlinks=true]{hyperref}

\newcommand{\fb}{f_\mathrm{b}}

\newcommand{\BPRP}{ {G_\mathrm{BP}-G_\mathrm{RP} }}

\newcommand{\Gbp}{G_\mathrm{BP}}
\newcommand{\Grp}{G_\mathrm{RP}}

\shorttitle{MiMO-CMD}
\shortauthors{Li \& Shao et al.}
\begin{document}
\begin{CJK*}{UTF8}{gbsn}

\title{MiMO: Mixture Model for Open Clusters in Color-Magnitude Diagrams}
\correspondingauthor{Zhengyi Shao, Lu Li}
\email{zyshao@shao.ac.cn, lilu@shao.ac.cn}

\author[0000-0002-0880-3380]{Lu Li (李璐)}
\affil{Key Laboratory for Research in Galaxies and Cosmology, Shanghai Astronomical Observatory, Chinese Academy of Sciences, 80 Nandan Road, Shanghai 200030, China.}
\affil{University of Chinese Academy of Sciences, No. 19A Yuquan Road, Beijing 100049, People’s Republic of China}
\affil{Centre for Astrophysics and Planetary Science, Racah Institute of Physics, The Hebrew University, Jerusalem, 91904, Israel}

\author[0000-0001-8611-2465]{Zhengyi Shao (邵正义)}
\affil{Key Laboratory for Research in Galaxies and Cosmology, Shanghai Astronomical Observatory, Chinese Academy of Sciences, 80 Nandan Road, Shanghai 200030, China.}
\affil{Key Lab for Astrophysics, Shanghai 200234, China}

%==============================================================

\begin{abstract}
We propose a mixture model of open clusters (OCs) in the color-magnitude diagrams (CMDs) to measure the OC properties, including isochrone parameters (age, distance, metallicity, and dust extinction), stellar mass function (MF), and binary parameters (binary fraction and mass-ratio distribution), with high precision and reliability. The model treats an OC in the CMD as a mixture of single and binary member stars and field stars in the same region. The cluster members are modeled using a theoretical stellar model, MF and binary properties. The field component is modeled nonparametrically using a separate field-star sample in the vicinity of the cluster. Unlike conventional methods that rely on stringent member selection, ours allows us to use a sample of more complete cluster members and attendant field stars. The larger star sample reduces the statistical error and diminishes the potential bias by retaining more stars that are crucial for age estimation and MF measurement. After validating the method with 1000 mock clusters, we measured the parameters of 10 real OCs using Gaia EDR3 data. The best-fit isochrones are consistent with previous measurements in general but with more precise age estimates for several OCs.The inferred MF slope is $-2.7$ to $-1.6$ for clusters younger than 2 Gyr, while older clusters appear to have significantly flatter MFs. The binary fraction is 30\% to 50\%. The photometric and astrometric distances agree well.

\end{abstract}


%==============================================================
\section{Introduction}
\label{sec:intro}

\defcitealias{Li2020d}{Li20}
\defcitealias{Dias2021a}{D21}

Galactic open clusters (OCs) are gravitationally bound stellar systems containing hundreds to tens of thousands of stars. They can be used as probes for a variety of astrophysical phenomena. First, it has been widely accepted that stars form in clusters or associations rather than in isolation. The clusters then evaporate and disrupt over time, enriching the field population \citep{Lada2003}. This makes OCs unique laboratories for investigating the formation, evolution, and dynamics of stars. Second, OCs exhibit a wide range of properties and are found at all ages and almost all locations in the Galactic disk, making them good tracers of disk structure and evolution \citep{Becker1970,Janes1982,Cantat-Gaudin2020b,Monteiro2021a}. All of this research depends on the precise measurement of OC properties.

The color-magnitude diagram (CMD) is used as a fundamental diagnostic tool for extracting the physical parameters of OCs. Because the stars in an OC formed at the same time in the same molecular cloud, they share the same age, metallicity, and distance. The single stars in this coeval population occupy a curve in a CMD. Comparison with theoretical isochrones hence allows a determination of OC properties.

In early studies, estimating the parameters has mostly relied on a by-eye comparison of the observational CMD with theoretical isochrones. This method depends significantly on subjective judgment, which prevents accurate measurements of parameters.

The first attempt beyond the "by-eye" approach was made by \citet[see also \citealt{Luri1992}]{Flannery1982a}, who searched for the best-fit isochrone by minimizing the sum of the distance between observation data points and their nearest points on the isochrone. Alternatively, \citet{Holland1992} alternatively considered the curve-to-curve distance between the isochrone and the single-star sequence extracted from the observation. Later, the curve-fitting approaches were further developed into a more probabilistic analysis \citep{Luri1992}. \citet{Hernandez2008a} employed a similar method but took a weighted likelihood function based on a given stellar mass function (MF). All of the above curve-fitting approaches only concern the geometrical location of the isochrone in CMD, which makes it difficult to incorporate observational errors and thus difficult to derive statistically meaningful uncertainties for parameter estimates.

Moreover, these curve-fitting techniques implicitly assume that the population is composed entirely of single stars, while unresolved binaries generally make up a significant fraction of OC photometric samples in real observation. Their existence broadens the main sequence: Equal-mass binaries appear 0.75 mag brighter than single stars and unequal-mass binaries lie between the two (see Figure 1 in \citealt{Li2020d} for illustration). Clearly, unresolved binaries must be taken into account to obtain precise estimates of OC parameters.

\citet{Tolstoy1996} proposed a solution using synthetic clusters with a binary population \citep[see also][]{Monteiro2010,Perren2015}. Given isochrone and binary parameters, they first created simulated observation data in the CMD and then estimated the goodness of fit of the parameters using the distances between the simulated data points and real observation. More advanced approaches model the CMD as a 2D probability density distribution \citep{Naylor2006,vonHippel2006a,Jeffery2016}. One advantage of these approaches is the ability to incorporate both the MF and unresolved binaries, including the fraction and mass-ratio distribution of binaries. However, these works confined the analysis to isochrone parameters and treated the rest as nuisance parameters. 

In this work, we present the first work that fits all cluster parameters simultaneously in a fully probabilistic and self-consistent way.

Another long-existing major problem in determining OC properties is the field-star contamination from foreground and background. Conventionally, most studies on cluster parameters rely on disentangling the cluster members from field stars in photometric and/or kinematic space explicitly.

In photometric space (CMD), the simplest approach might be rejecting stars far from the cluster main sequence beyond some arbitrary thresholds \citep{Claria1986,Tadross2001,Roberts2010}. Another way is by randomly rejecting field stars based on the membership probability estimated from the difference in stellar density between the cluster region and adjacent field region of comparable size in the CMD\citep{Baade1983a,Bonatto2007,Maia2010b}. The kinematics (especially proper motions) can also serve as a powerful membership diagnosis. \citet{Vasilevskis1958} and \citet{Sanders1971} estimated the membership probabilities by modeling the cluster and field stars as bivariate Gaussian distributions separately in the vector point diagram. This approach has been adopted and improved in numerous subsequent studies \citep{Zhao1990,Balaguer-Nunez2004,Krone-Martins2004,Krone-Martins2014a,Sarro2014,Pera2021a}.

In recent years, the European Space Agency Gaia mission along with its second and early third data releases (\citealt{Collaboration2018a,GaiaCollaboration2020c}, hereafter DR2 and EDR3 respectively), has represented the deepest all-sky astrometric and photometric survey ever conducted. It allows researchers to establish more accurate membership and thus more reliable estimation of OC properties \citep[hereafter D21]{Castro-Ginard2018a,Cantat-Gaudin2018b,Bossini2019a,Monteiro2020a,Dias2021a}. Nevertheless, field stars still exist. The problem is particularly severe for OCs projected against the disk. 

It is worth pointing out an intrinsic dilemma existing in all approaches based on membership determination: A loose sample selection criterion cannot avoid field-star contamination, while a stringent criterion may remove too many cluster members. For the latter, it is not only a problem of losing statistical precision. As we will show in Section \ref{sec:compare_d21}, the stars near the main-sequence turnoff region are crucial in determining the cluster age; their absence due to a stringent criterion may result in an overestimation of the age. Therefore, the best way is not to be afraid of field stars and clean them out. Rather, we can acknowledge the presence of field-star contamination in the fitting sample and incorporate them as a model component.

In this work, we present the MiMO (MIxture Model of Open clusters), a Bayesian framework that models the CMD of an observed OC as a mixture distribution of single stars, unresolved binaries, and field stars.

Based on theoretical isochrone models, the single-stellar population is characterized by age, distance, metallicity, extinction, and mass function, while the binary population is characterized by the binary fraction and mass-ratio distribution function. The field-star population is made up of a wide range of stellar populations at different distances, and it changes from cluster to cluster. For simplicity, we construct the field-star distribution nonparametrically through an adaptive smoothing on the field stars in the same sky region but distinguish them proper motion. Here we do not distinguish the cluster members from the field in advance. As mentioned above, this not only has the statistical advantage of sample size but also diminishes the bias in age by retaining more bright stars. The comprehensive physical models and advanced statistics enable the determination of all the OC properties simultaneously with high precision and thus permit fruitful discoveries.

As a demonstration, we then apply the method to 10 sample OCs with Gaia EDR3 photometric data \citep{Riello2021} in the CMD of ($G_\mathrm{BP}-G_\mathrm{RP}$, $G$). By rigorously incorporating field stars, our method can exploit the precise photometry brought by EDR3 and extract the physical parameters of OCs from their CMD. 

This paper is organized as follows: In the next section, we introduce the mixture model and define the likelihood function. We validate the method with synthetic clusters in Section \ref{sec:Mock} and apply MiMO to 10 real OCs in Section \ref{sec:real_OC}. We compare the results with the literature and discuss the advantages and limitations of MiMO in Section \ref{sec:disscuss}. We conclude in Section \ref{sec:Conclusion}.

%==============================================================
\section{Method}
\label{sec:MM}

The mixture model has been widely adopted in astronomy (see \citealt{Kuhn2017a} for a review). For example, in stellar cluster studies, the mixture model has been used to estimate the membership probability in kinematic data space \citep{Zhao1990}.

\citet[hereafter \citetalias{Li2020d}]{Li2020d} modeled the member stars in CMDs as a mixture of single and unresolved binary stars, then measured the binary fraction and binary mass-ratio distribution. In this work, we extend the model of \citetalias{Li2020d} by further including the field component. This allows us to measure all the parameters that shape the distribution in CMD and estimate the fraction of field contamination simultaneously.

We present the method in the following subsections. The details of the numerical implementation are provided in Appendix \ref{sec:implementation}.

\subsection{Model of cluster members}
\label{sec:MM:cl}

Following \citetalias{Li2020d}, we model the member stars as a mixture of single stars and unresolved binary populations. Here we provide a brief introduction and refer interested readers to \citetalias{Li2020d} (section 2.1 therein) for details.

We denote the probability density distribution of member stars in the space of apparent magnitude, $m$, and color, $c$, by $\phi_\mathrm{cl}(m,c\ |\ \Theta_\mathrm{cl})$, where $\Theta_\mathrm{cl}$ is the ensemble of cluster properties, including parameters that characterize the isochrone, MF, and binaries (see Table \ref{tab:model-paras}). In this work, we use the PARSEC theoretical isochrones \citep{Bressan2012b}\footnote{PARSEC version 1.2S, \url{http://stev.oapd.inaf.it/cgi-bin/cmd}} and Gaia EDR3 photometric system (\citealt{Riello2021}, $G$ and $\BPRP$). Considering the wide photometric filter bands of Gaia, the delta bolometric correction ($\Delta BC = BC_{A_V} - BC_{A_V =0}$) for a given spectrum is not a linear function of $A_V$ (see Figure 4 in \citealt{Chen2019c}). Therefore, we use the variable extinction model YBC \citep{Chen2019c}\footnote{\url{http://stev.oapd.inaf.it/YBC}} instead of a constant extinction coefficient for all $A_V$s.

For a theoretical isochrone characterized by age and metallicity,
we convert the absolute magnitude and the intrinsic color to the apparent magnitude and the reddened color $(m, c)$ according to the distance module and dust extinction.
We emphasize that MiMO, as a general method, can naturally apply to
any other stellar evolution models, photometric bands, or extinction models as well.

We omit $\Theta_\mathrm{cl}$ in the following equations for simplicity. Combining single stars and binaries, we have
\begin{align}\label{eq:phi-cl}
  \phi_\mathrm{cl}(m,c) = & (1-f_{\rm b}) \phi_\mathrm{s}(m,c) + f_{\rm b} \phi_{\rm b}(m,c),
\end{align}
where $\phi_\mathrm{s}$ and $\phi_\mathrm{b}$ are the model number density of single stars and binary stars, respectively, and $f_\mathrm{b}$ is the fraction of binary stars. 

Given an isochrone, the distribution of a single star in the CMD, $\rho_\mathrm{s}$, is a $\delta$ function of its mass $\mathcal{M}$. Similarly, the distribution of a binary, $\rho_\mathrm{b}$, is also a $\delta$ function of $(\mathcal{M},q)$, where $q=\mathcal{M}_2/\mathcal{M}_1$ is the mass ratio between the two components in a binary  ($\mathcal{M}_1>\mathcal{M}_2$). Then, the distribution of a stellar population is,
% --
\begin{equation}\label{eq:phi-s}
\phi_\mathrm{s}(m,c)= \int_{\mathcal{M}} \rho_{\rm s}(m, c |\mathcal{M} )\mathcal{F}_\mathrm{MF}(\mathcal{M})d\mathcal{M}. \\
\end{equation}
% --
and
\begin{equation}\label{eq:phi-b}
\phi_\mathrm{b}(m,c)= \int_{\mathcal{M}} \int_{q} \rho_{\rm b} (m, c |\mathcal{M},q)\mathcal{F}_\mathrm{MF}(\mathcal{M})
\mathcal{F}_q(q) d \mathcal{M} dq. \\
\end{equation}
where $\mathcal{F}_\mathrm{MF}$ is the mass function, and $\mathcal{F}_q(q)$ is the binary mass-ratio distribution.
Following \citetalias{Li2020d},
we define $\fb$ as the fraction of binaries with $q>0.2$ among the member stars,
because binaries with lower $q$ are nearly indistinguishable from single stars.

\begin{table}[htbp]
{\centering
\caption{Description and the prior range of parameters in MiMO}
\label{tab:model-paras}
\begin{tabular*}{1\columnwidth}{l @{\extracolsep{\fill}} ll}
	\hline
	\hline
	 & Range &Description \\
	\midrule
	\multicolumn{3}{l}{\it{Isochrone Parameters}} \\
	\cmidrule(l){1-3}
	\quad logAge & $[6.2, 10.1]$& $\log_{10}$ cluster age (year)\\
	\quad  $\mathrm{DM}$ & $[3, 15]^*$ & distance modulus (mag)  \\
	\quad  $A_V$ & $[0, 3]^*$& dust extinction in the $V$ band (mag) \\
	\quad  $\mathrm{[Fe/H]}$ & $[-2.1, 0.5]$ & $\log_{10}$ iron-to-hydrogen ratio\\
	 & &   relative to the Sun (dex) \\
	\midrule
	\multicolumn{3}{l}{\it{Mass Function Parameter}} \\
	\cmidrule(l){1-3}
	\quad  $\alpha_\mathrm{MF}$ & $[-4, 2]$& power-law index of Salpeter's MF \\
	\midrule
	\multicolumn{3}{l}{\it{Binary Parameters}} \\
	\cmidrule(l){1-3}
	\quad  $f_\mathrm{b}$& $[0,1]$ & fraction of binaries with $q>0.2$ \\
	& & among the member stars\\
	\quad  $\gamma_{q}$ & $[-2, 2]$& power-law index of the binary mass \\
	  & &  ratio distribution \\
	\midrule
	\multicolumn{3}{l}{\it{Field Parameter}} \\
	\cmidrule(l){1-3}
	\quad  $f_\mathrm{fs}$ & $[0, 1]$& fraction of field stars in the sample \\
	\hline
	\hline
\end{tabular*}}
\tablecomments{* The distance modulus range corresponds to a distance of 40 pc to 10 kpc. The above ranges of the distance modulus and extinction are chosen specifically for the clusters used in this work.
It is possible to use wider ranges when necessary.}
\end{table}

\noindent\textbf{The forms of $\mathcal{F}_\mathrm{MF}$ and $\mathcal{F}_q$ }

Following \citetalias{Li2020d}, here we adopt a single power-law MF (\citealt{Salpeter1955}) for simplicity,
\begin{equation}\label{eq:F-MF}
\mathcal{F}_\mathrm{MF} (\mathcal{M}) = \frac{dN}{d\mathcal{M}} \propto \mathcal{M}^{\alpha_{\rm MF}},
\end{equation}
where the power-law index $\alpha_{\rm MF}$ is a parameter to be fit. For observation, other forms of MFs have also been proposed, e.g., a tapped power law \citep{2002Sci...295...82K} or a log-normal \citep{2003PASP..115..763C}. Nevertheless, the fitted power-law index $\alpha_{\rm MF}$ can still reflect the average trend of the true MF. In addition, we find that the inferred isochrone parameters are not sensitive to the assumed form of the MF (see Section \ref{sec:mf} for more discussions).

The binary mass ratio is usually assumed to follow a power-law distribution as well (\citealt{Kouwenhoven2007a,Duchene2013,Reggiani2013}),
%--
\begin{equation}\label{eq:F-q}
\mathcal{F}_q (q)= \frac{dN}{dq} \propto q^{\gamma_q},
\end{equation}
%--
where the power-law index $\gamma_q$ is the parameter that shapes the mass-ratio distribution.

Finally, it is worth pointing out that it is possible to use any other forms of the MF and $q$ distribution depending on the specific cluster of concern.

\subsection{Model of field stars}
\label{sec:field model}
Besides the member stars, a cluster region also contains foreground and background stars. These field stars belong to diverse stellar populations at different distances. Naturally, the field-star distribution varies from cluster to cluster in the Milky Way, and hence modeling it theoretically remains impossible without a more thorough knowledge of the Milky Way. Instead, assuming that the field stars in the cluster region follow the same distribution in the CMD as those in the neighboring sky area, one can construct a realistic model of field stars empirically from the adjacent sky area. Similarly, one can also use the field stars in the cluster region but with different kinematics from the cluster members. 
We find that selecting field stars by sky area or by kinematics seems to make little difference in general.
We adopt the second approach in this work for simplicity in sample selection.

For each target cluster, we first select a field-star sample in the same sky area as the cluster but distinguished from the cluster in the proper-motion space, as detailed in Section \ref{sec:data_select}. Then, the expected probability density distribution of field stars in the CMD is given by the kernel density estimation technique,
\begin{align}
    \phi_\mathrm{fs}(m,c) = \frac{1}{N_\mathrm{fs}} \sum_{i=1}^{N_\mathrm{fs}} \mathcal{N}(m|m_i,\epsilon_m)\mathcal{N}(c|c_i, \epsilon_c),
    \label{eq:field stars_prob}
\end{align}
where $N_\mathrm{fs}$ is the sample size of field stars, $\mathcal{N}$ is the Gaussian smoothing kernel, $m_i$ and $c_i$ refer to the $i$th star in the sample, and $\epsilon_m$ and $\epsilon_c$ are the smoothing length in each dimension. Considering that a constant smoothing size may lead to oversmoothing in the crowded region and undersmoothing elsewhere, we choose the smoothing size adaptively \citep[following][]{Li2019d} so that a field star with a low local density in the CMD has a greater smoothing size. 

We emphasize that it is not obligatory to select the field-star sample by kinematics. In the case of pure photometric data,
one may instead take stars in the neighboring sky area outside the cluster (e.g., a ring between 3$r_{50}$ and 5$r_{50}$) to construct the field model.

\subsection{Mixture model}
\label{sec:method:MM:MM}

The mixture distribution of cluster members and field stars in the CMD is then written as
\begin{align}
    \phi(m,c\ |\ \Theta) = (1-f_\mathrm{fs}) \phi_\mathrm{cl}(m,c\ |\ \Theta_\mathrm{cl}) + f_\mathrm{fs}\phi_\mathrm{fs}(m,c),
    \label{eqn:phitot}
\end{align}
where $f_\mathrm{fs}$ is the fraction of field stars in the fitting sample,
and $\Theta$ is an ensemble of $\Theta_\mathrm{cl}$ and $f_\mathrm{fs}$.

\subsection{Likelihood}
\label{sec:method:MM:LH}

We can further add observational errors and sample selection into the likelihood. Given the observational errors in magnitude and color, $\sigma_{m}$ and $\sigma_{c}$, the probabilities of this star belonging to the cluster or the field ($X=\mathrm{cl}$ or fs) are expected to be the convolution of the model and measurement uncertainties,
%----
\begin{multline}\label{eq:obs_psi}
    \psi_{X}(m,c\,|\,\sigma_{m},\sigma_{c}) = \mathcal{C}_X\int_{m',c'} \phi_{X}(m',c')  \\
           \cdot \mathcal{N}(m\,|\,m',\sigma_{m})\mathcal{N}(c\,|\,c',\sigma_{c})\,dm'dc',
\end{multline}
%--
where $\mathcal{C}_X$ is a normalization factor for each component. It is noteworthy that each star follows a different probability distribution depending on its own observational error.

If the observation data is truncated by flux limits, say $m_1<m<m_2$, then $\mathcal{C}_X$ should be taken to satisfy
\begin{equation}\label{eq:C}
     \int_{m_1}^{m_2}\int_c \psi_{X}(m,c) \,dm dc =1,
\end{equation}
see \citetalias{Li2020d} for details. In this work, we select stars in the magnitude range of from $G=18$ to the brightest star in each cluster region.

Similar to Equation (\ref{eqn:phitot}), the total probability density in the presence of observational error is given by
\begin{align}
    \psi(m,c|\ \Theta) = (1-f_\mathrm{fs}) \psi_\mathrm{cl}(m,c\,|\ \Theta_\mathrm{cl}) + f_\mathrm{fs}\psi_\mathrm{fs}(m,c).
    \label{eqn:lik_err}
\end{align}

Putting it together, the likelihood (probability) of observing a sample of $N$ stars, $\{m_i,c_i\}_{i=1, \ldots, N}$, under a given model is
\begin{equation}\label{eq:lhtot}
    \mathcal{L}(\{m_i, c_i\}|\,\Theta) = \prod_{i=1}^{N}\ \psi(m_i, c_i|\,\Theta),
\end{equation}
where $\Theta$ is the ensemble of all the parameters involved as listed in Table \ref{tab:model-paras}. 

\subsection{Bayesian inference}
\label{sec:method:MM:parameters}

According to the Bayesian formula, we can infer the posterior probability distribution of $\Theta$,
\begin{equation}\label{eq:pdf}
    \mathcal{P} (\Theta|\,\{m_i, c_i\}) \propto \mathcal{L}(\{m_i, c_i\}|\,\Theta)\cdot \pi(\Theta),
\end{equation}
where $\pi(\Theta)$ represents our prior knowledge of the parameters. 

The range of each parameter used in the fitting is shown in Table \ref{tab:model-paras}. One may take flat distributions or previous independent measurements as priors. In this work, for real OCs, we use the spectroscopy measurements in the literature as the prior for metallicity, a fixed value for $\gamma_q$,  and flat priors for the rest of the parameters.

The posterior distribution of the parameters can be obtained through sampling methods.
Specifically, in this work we employ the nested sampling method (\citealt{Skilling2004a,Skilling2006})
implemented by the public package \texttt{dynesty} \citep{Speagle2020}%
\footnote{\url{https://github.com/joshspeagle/dynesty}}
to obtain the posterior distribution of the parameters (see Appendix \ref{sec:dynesty} for more details.)
We then compute the marginal distribution for each parameter. The fitted parameters and their statistical errors (i.e.\ formal errors) are presented in terms of the median value and half of the [16\%, 84\%] interval, respectively, throughout the paper.

Finally, note that we have ignored several effects, e.g., the stellar rotation and the intrinsic variation in the dust extinction, which can broaden the main sequence and turnoff region. We believe that these effects are secondary, and our inferred parameters should present an average performance over these effects. Nevertheless, it could be straightforward to include such effects in our statistical framework, once corresponding models become available in the future. 

\subsection{Membership probability}
Once the best-fit parameters are found, we can derive the membership probability for each sample star as a natural byproduct,
\begin{align}\label{eq:pmem}
   p_\mathrm{memb} =\frac{ (1-f_\mathrm{fs}) \psi_\mathrm{cl}(m,c\,|\,\Theta_\mathrm{cl}) }{ \psi(m,c\,|\,\Theta)}.
\end{align}

Similarly, one can further derive the probability of a single star or a binary.

%==============================================================
\section{Validation with mock clusters}
\label{sec:Mock}

Before applying the mixture model to real observation data, we test its validity and accuracy of our mixture model with mock clusters.

\subsection{Generating mock samples}

%----
 \begin{figure*}[!htbp]
  \centering
  \includegraphics[width=1\textwidth]{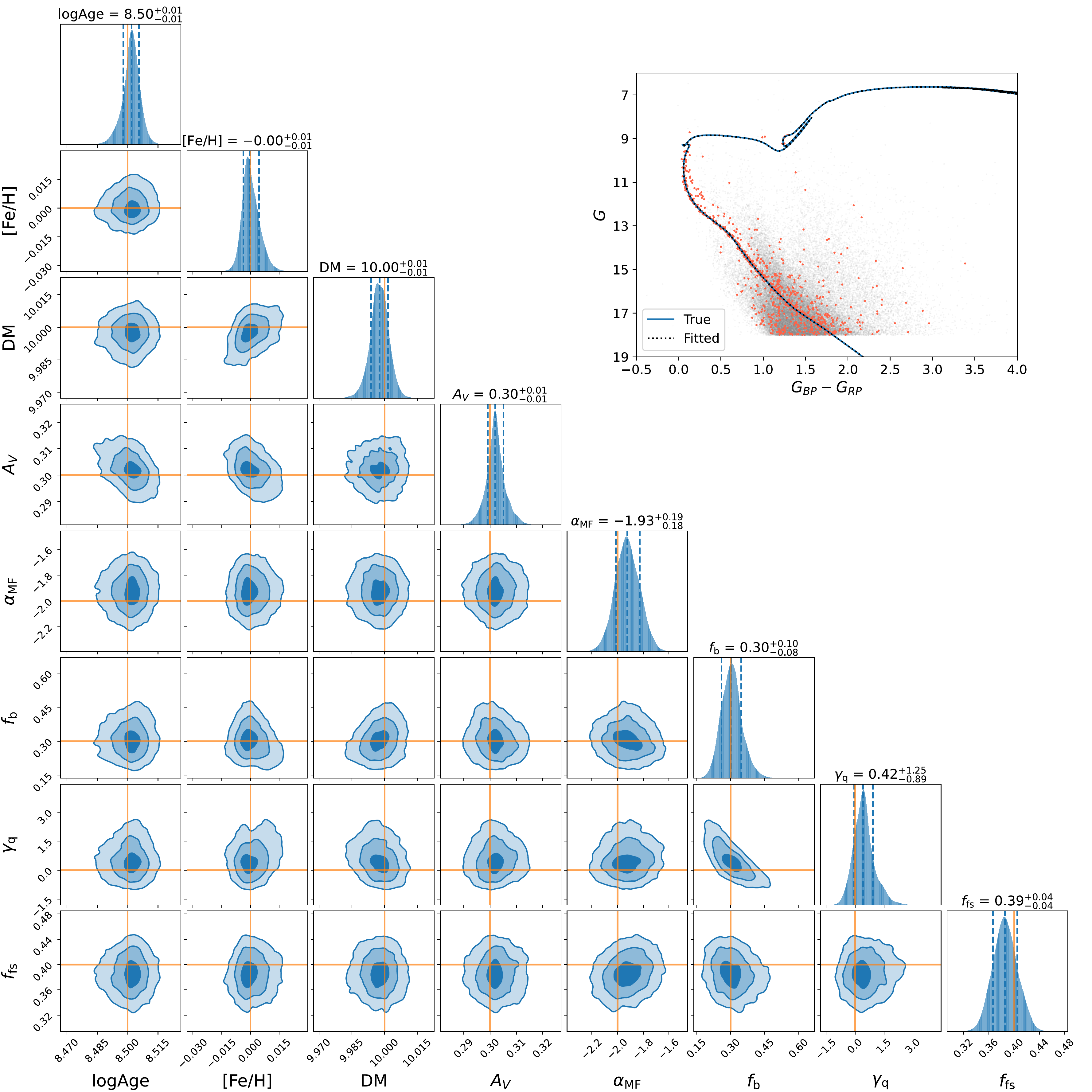}
  \caption{
  \textit{Left:} the probability density distributions of parameters based on the nested Sampling for an example mock cluster. The blue contours correspond to the $1\sigma$, $2\sigma$ and $3\sigma$ (39.3\%, 86.5\% and 98.9\%) confident levels, whereas the orange lines indicate the true values. Histograms show the marginalized probability distributions for estimated values of each parameter. Blue dashed lines indicate the 15.9th, 50.0th, and 84.1th percentiles (See also the numbers above each panel.). \textit{Upper right:} The CMD of this example mock OC. The fitting sample including both member and field stars are shown as red dots; the field sample that used for building the field model is shown as grey dots. The blue curve and black dashed line represent the true isochrone and fitted isochrone, respectively.}
  \label{fig:mock_single}
\end{figure*}
%----

%----
 \begin{figure*}[!htbp]
  \centering
  \includegraphics[width=1\textwidth]{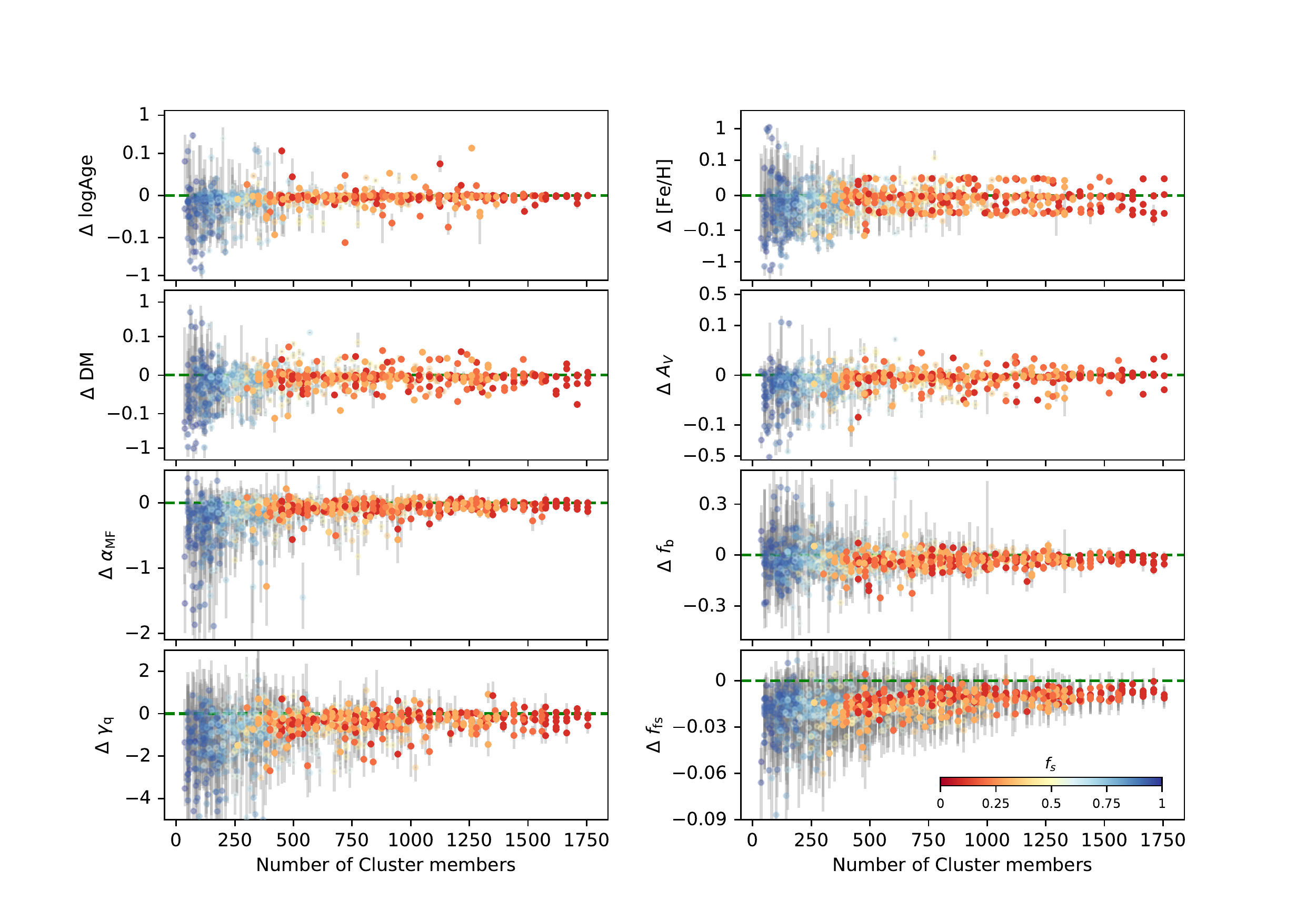}
  \caption{Residual between the best-fit and the true value of each parameter as a function of the number of member stars. Each circle represents a mock cluster, color-coded by the $f_\mathrm{fs}$ of this cluster. In order to better show the details, we show the $y$-axis of the upper four panels (logAge, DM, $\mathrm{[Fe/H]}$, and $A_V$) in a ``symmetrical log'' scale, which uses a linear scale in the range $[-0.1, 0.1]$ and a logarithmic scale otherwise.}
  \label{fig:mock_res_all}
\end{figure*}
%----

We generate mock clusters following the procedure proposed by \citetalias{Li2020d}. For a mock cluster with given parameters, we first randomly assign a mass $\mathcal{M}$ to each star according to the MF (Equation \ref{eq:F-MF}). Then, a fraction $f_{\rm b}$ of them are treated as binaries. For each binary, we assign a companion secondary star of mass $q \mathcal{M}$, where $q$ is chosen randomly from the mass-ratio distribution  (Equation \ref{eq:F-q}).

Subsequently, we derive the magnitude, $G$, and color, $G_\mathrm{BP}-G_\mathrm{RP}$, of each single star or binary according to the PARSEC theoretical isochrone \citep{Bressan2012b} and Gaia EDR3 photometric system \citep{Riello2021}. Given the cluster age,  metallicity, and extinction, the isochrone is downloaded from the PARSEC web interface with the variable extinction model \citep{Chen2019c} integrated.
We then convert the absolute magnitude to the apparent magnitude using the distance module of the mock cluster. 

To mimic realistic observations, we add random noise according to the magnitude-dependent observational errors in each photometric band \citep{Riello2021} along with an additional scatter of 0.01 mag (\citetalias{Li2020d}). The additional scatter is motivated by the fact that the actual dispersion in observed OCs is significantly larger than the Gaia formal error as reported by \citetalias{Li2020d}.

We then add a field-star component to the mock cluster. As mentioned in Section \ref{sec:field model}, it is difficult to simulate a field component with theoretical models because field stars contain complex stellar populations as well as complex distance distributions. As a practical operation, we randomly selected 50,000 stars from the Gaia EDR3 with $G<18\ \mathrm{mag}$ from the sky region within $5^\circ$ of the antigalactic direction as the field-star database.

We generated 1000 mock OCs in CMD with a magnitude limit of $G<18\ \mathrm{mag}$.  For each cluster, its parameters are chosen randomly from the parameter range listed in Table \ref{tab:model-paras}, to cover a wide range of cases to test the general applicability of our method. The total number of stars, $N_\mathrm{tot}$, is randomly selected from 500 to 2000. $N_\mathrm{cl}=(1-f_\mathrm{fs}) N_\mathrm{tot}$ member stars are generated by the cluster model as mentioned above, and $N_\mathrm{fs}= f_\mathrm{fs} N_\mathrm{tot}$ field stars are randomly selected from the field-star database.

\subsection{Mock results}
\label{sec:mock_res}

We use MiMO with flat priors to estimate the properties of the mock clusters. For each mock OC, 5000 stars are randomly chosen from the field-star database to construct the field-star model.

As the first example, we start with a single mock cluster of 1000 stars with $f_\mathrm{fs}=0.4$ (i.e., 600 member stars and 400 field stars) and typical parameters. The CMD of this mock OC is shown in the upper-right panel of Figure \ref{fig:mock_single}. Similar to real OCs in observations (e.g., Figure \ref{fig:res_cmd}), one can clearly recognize the main sequence, binary sequence, and field contamination.

The posterior distribution of the parameters is shown in the lower-left panels of Figure \ref{fig:mock_single}. All the inferred parameters are in good agreement with their true values. The uncertainties are small, especially for the isochrone parameters (logAge, [Fe/H], DM, $A_V$). As shown in the upper-right panel, the best-fit isochrone matches the data and the true isochrone perfectly, despite the field-star contamination being as high as 40\% in the input sample.

Looking at the parameter distribution in Figure \ref{fig:mock_single}, we see an anticorrelation between the two binary parameters, $f_\mathrm{b}$ and $\gamma_q$. As discussed in \citetalias{Li2020d}, this degeneracy is because it is hard to distinguish the low-mass-ratio binaries from single stars due to the observational error. It is also worth noting that the binary parameters are largely independent of the other OC parameters (isochrone, MF, and field fraction), which thus allows one to treat them separately when necessary. For example, \citetalias{Li2020d} focused on estimating binary parameters with all other parameters fixed.

We also see anticorrelations between $A_V$ and metallicity, and $A_V$ and age, and a positive correlation between distance and metallicity. This is expected from the way each parameter affects the shape and location of isochrones. These correlations would be even more prominent when the statistical uncertainties are larger, due to, e.g., larger observational errors or a smaller sample size.

The parameter degeneracies are undesirable in some cases. In practice, the imperfection in the stellar model and the broadening of the main sequence (e.g., due to inhomogeneous internal extinction and stellar rotation; see \citetalias{Li2020d} for more discussion) may bias the estimate of certain parameters. Such bias will then affect the rest of the parameters through degeneracies. Fortunately, this issue can be greatly alleviated by introducing prior information from independent measurements. The natural and rigorous treatment of priors is a unique advantage of the Bayesian method.

\begin{table}%[htbp]
\caption{Typical internal precision of MiMO for the mock clusters.}
\centering
\begin{tabular*}{0.8\columnwidth}{@{\extracolsep{\fill}} lll}
	\hline
	\hline
	 Parameter & Statistical Error & Actual Error \\
	\midrule
	logAge & 0.01	 & 0.018 \quad[dex]\\
	$\mathrm{[Fe/H]}$ & 0.018	 & 0.031  \quad [dex]\\
	DM & 0.012	 & 0.022  \quad [mag]\\
	$A_V$ & 0.0091	 & 0.016  \quad [mag]\\
	$\alpha_\mathrm{MF}$ & 0.1	 & 0.098 \\
	$f_\mathrm{b}$ & 0.056	 & 0.051 \\
	$\gamma_\mathrm{q}$ & 0.51	 & 0.52 \\
	$f_\mathrm{fs}$ & 0.014	 & 0.0065 \\
	\hline
	\hline
\end{tabular*}
\label{tab:paras-accuracy}
\end{table}

We then apply our method to all 1000 mock OCs. The differences between the inferred parameters and their true values are shown in Figure \ref{fig:mock_res_all}. Except for a small bias of $\sim\! 1\%$ in the field-star fraction, all the cluster parameters are unbiased overall. As expected, the error of each parameter is smaller for clusters with a larger number of member stars $N_\mathrm{cl}$ or lower field contamination $f_\mathrm{fs}$. When $N_\mathrm{cl}$ is large, [Fe/H] residuals show discrete features at $\pm 0.05$ dex which reveals the grid-based nature of our model implementation (see Appendix \ref{sec:implementation} for details).
We find that the precision is nearly independent of the isochrone and MF parameters. When the field contamination is controlled to a reasonable level ($<50\%$), the errors are insensitive to $f_\mathrm{fs}$ as well. 

In Table \ref{tab:paras-accuracy}, we list the average statistical error (the so-called formal error), $({\sum_i\sigma_i^2/n})^{1/2}$, of each parameter for mock clusters with $f_\mathrm{fs}<50\%$,
in comparison to the actual mean deviation from the truth $({\sum_i (X_{\mathrm{fit},i}-X_{\mathrm{true},i})^2/n})^{1/2}$.
They present the typical precision of our method when using the high-precision photometric data from Gaia EDR3 (ignoring the systematics due to the possible mismatch between isochrones and observation). For clusters of different sample sizes, the precision roughly scales as $\sqrt{800/N_\mathrm{memb}}$. Ideally, the two uncertainties in Table \ref{tab:paras-accuracy} are expected to be equal, as in the cases for the MF and binary parameters.
However, the statistical error underestimates the true deviation for the isochrone parameters by a factor of $\sim\! 2$. It is probably because the grid-based model implementation sets a lower limit of the true precision, thus we cannot faithfully resolve the posterior distribution at scales significantly smaller than the grid resolution (0.05dex for logAge and [Fe/H] grids in this paper; see Appendix \ref{sec:implementation} for details).
One thus should be careful when interpreting very small statistical errors.

Here we report a precision of about 0.02 dex for logAge, 0.02 mag for $A_V$, and $<0.01$ kpc for the distance of a cluster located within 2 kpc. For comparison, Asteca \citep{Perren2015} with SDSS data reported a precision of 0.16--0.36 dex for logAge, 0.2 mag for $A_V$, and 0.2--1 kpc for distance, and \citet{Dias2021a} with Gaia DR2 data reported a precision of 0.1 dex for logAge, 0.1 mag for  $A_V$, and 0.1 kpc for distance of typical OCs. Therefore, the statistical error of OC parameters has been significantly reduced due to our advanced method and high-precision data from Gaia EDR3. Finally, it is worth emphasizing that the above precision is about the statistical error in the context of the methodology. When applying to observations, the imperfectness of stellar models and unconsidered observational effects (e.g., variation in the cluster internal extinction, flawed calibration of photometric system) inevitably introduces additional systematical errors, which can be much larger than the statistical error. 

%==============================================================
\section{Application to observed clusters}
\label{sec:real_OC}

In this section, we apply MiMO to infer the parameters of real OCs. To demonstrate MiMO's capacity in different situations, we select 10 OCs (listed in Table \ref{table:keni_paras}) with diverse ages, distances, and field-star contamination levels. As discussed in Section \ref{sec:mock_res}, using the [Fe/H] from spectroscopic measurements as a prior can break degeneracies. Thus, the selected OCs are required to have at least one [Fe/H] measurement from high-resolution spectroscopy \citep{Netopil2016a,Carrera2019a,Donor2020,2021MNRAS.503.3279S}. 

\subsection{Fitting sample selection}
\label{sec:data_select}

For each cluster, we select the fitting sample from the Gaia EDR3 source catalog \citep{GaiaCollaboration2020c}. Unlike conventional methods that rely on a "pure" member sample from stringent filters, MiMO can naturally handle the field-star contamination with a mixture model. It thus allows us to exploit a sample containing more complete member stars but (inevitably) accompanied by moderate field-star contamination. As will be shown later in Section \ref{sec:compare_d21} and Section \ref{sec:mf}, the high completeness of member stars enabled by MiMO is crucial for accurate  age estimation and reliable MF measurement.

We start the sample selection with stars brighter than $G=18$ mag within $r_\mathrm{max}=3\,r_{50}$ from the cluster center in the sky, where $r_{50}$ is the radius expected to contain half of the members \citep{Cantat-Gaudin2020b}. The large aperture of $r_\mathrm{max}$ guarantees to a highly complete population of cluster members is encompassed.\footnote{This aperture is expected to enclose 99\% member stars for a typical OC ($r_t/r_c \simeq 2$) or 90\% member stars for an extremely loose OC ($r_t/r_c \simeq 20$), where the \citet{King1962} profile is assumed and the concentration parameter $r_t/r_c$ is defined as the ratio between the tidal radius and the core radius.
}

\begin{table*}
\begin{center}
	\caption{Basic astrometry properties of the 10 OCs used in this study.}
	\small\addtolength{\tabcolsep}{2pt}
	\begin{tabular}{lrrrrrrrrrr}
	\hline
	\hline

OC & $\alpha$ & $\delta$ & $r_{50}$ & $\mu_\alpha^\ast$ & $\sigma_{\mu_\alpha^\ast}$ & $\mu_{\delta}$ & $\sigma_{\mu_{\delta}}$ & $\varpi$ & $\sigma_{\varpi}$ & Distance\\
  & [deg] & [deg] & [deg] & [mas\,yr$^{-1}$] & [mas\,yr$^{-1}$] & [mas\,yr$^{-1}$] & [mas\,yr$^{-1}$] & [mas] & [mas] &[pc]\\
	\midrule

   ASCC~21 & 82.179 & 3.527 & 0.410 & 1.404 & 0.263 & $-0.632$ & 0.238 & 2.866 & 0.131 & 341\\
Trumpler~3 & 48.004 & 63.218 & 0.276 & $-3.354$ & 0.168 & $-0.112$ & 0.126 & 1.461 & 0.056 & 693\\
 Roslund~6 & 307.185 & 39.798 & 1.004 & 5.875 & 0.340 & 2.155 & 0.274 & 2.809 & 0.076 & 375\\
  NGC~2287 & 101.499 & $-20.716$ & 0.332 & $-4.339$ & 0.178 & $-1.381$ & 0.199 & 1.360 & 0.057 & 688\\
  NGC~2482 & 118.787 & $-24.263$ & 0.149 & $-4.722$ & 0.094 & 2.189 & 0.100 & 0.720 & 0.053 & 1285\\
  NGC~2447 & 116.141 & $-23.853$ & 0.202 & $-3.551$ & 0.158 & 5.068 & 0.167 & 0.968 & 0.055 & 1018\\
  NGC~2527 & 121.246 & -28.122 & 0.408 & $-5.549$ & 0.229 & 7.275 & 0.212 & 1.536 & 0.070 & 664\\
  NGC~2506 & 120.010 & $-10.773$ & 0.088 & $-2.571$ & 0.143 & 3.912 & 0.113 & 0.292 & 0.078 & 3191\\
  NGC~2682 & 132.846 & 11.814 & 0.166 & $-10.986$ & 0.193 & $-2.964$ & 0.201 & 1.135 & 0.051 & 889\\
   NGC~188 & 11.798 & 85.244 & 0.272 & $-2.307$ & 0.139 & $-0.960$ & 0.146 & 0.507 & 0.046 & 1698\\

	\hline
	\hline
	\end{tabular}
\label{table:keni_paras}
\end{center}
\tablecomments{Columns: R.A. ($\alpha$), decl ($\delta$), radius containing half of the member stars ($r_{50}$), 
mean and dispersion of cluster members' proper motions in R.A. and decl ($\mu_\alpha^\ast$, $\sigma_{\mu_\alpha^\ast}$, $\mu_{\delta}$, and $\sigma_{\mu_\delta}$), mean and dispersion of member parallax ($\varpi$ and $\sigma_{\varpi}$), and distance.
The astrometry parameters are measured by \citet{Cantat-Gaudin2020b} using the Gaia DR2 astrometric data.
% Note that the distance is provided for reference but not used in our fitting procedure.
}

\end{table*}

The OCs in the solar neighborhood can have a very large angular size, hence suffering more from the field contamination within $r_\mathrm{max}$. For nearby OCs closer than 500 pc (i.e., parallax $\varpi > 2$mas), we further remove the obvious background stars with $\varpi' < \varpi_\mathrm{cl}-6\sigma_{\varpi,\mathrm{cl}}$, where $\varpi_\mathrm{cl}$ and $\sigma_{\varpi,\mathrm{cl}}$ represent the average and dispersion of the cluster members' parallaxes \citep{Cantat-Gaudin2020b}.

In the proper-motion space, for OCs closer than 500 pc, we pick stars with $\Delta \mu < 6 \sigma_{\mu, \mathrm{cl}}$ as the fitting sample, where $\Delta \mu = \sqrt{(\Delta {\mu_\alpha^\ast})^2+(\Delta {\mu_\delta})^2}$ is the deviation from the cluster's average proper motion and $\sigma_{\mu, \mathrm{cl}}=\sqrt{\sigma^2_{\mu_\alpha^\ast}+\sigma^2_{\mu_\delta}}$ is the root sum squares of the cluster's dispersion in two directions. For an OC farther than 500 pc, the fitting sample is selected in a smaller aperture in proper motion, $\Delta \mu < 4 \sigma_{\mu, \mathrm{cl}}$.

We summarize the above sample selection for the reader's convenience,
\begin{compactitem}
\item for clusters with $\varpi_\mathrm{cl} > 2$mas:
\begin{compactitem}
    \item $G<18$ mag.,
	\item $r<3\,r_{50}$,
	\item $\varpi > \varpi_\mathrm{cl}-6\sigma_{\varpi,\mathrm{cl}}$,
	\item $\Delta \mu < 6 \sigma_{\mu, \mathrm{cl}}$;
\end{compactitem}
\item for clusters with $\varpi_\mathrm{cl} < 2$mas:
\begin{compactitem}
    \item $G<18$ mag.,
	\item $r<3\,r_{50}$,
	\item $\Delta \mu < 4 \sigma_{\mu, \mathrm{cl}}$;
\end{compactitem}
\end{compactitem}
where $r$ is the separation from the cluster center in the sky and $\varpi$ is the parallax of a star. One may note that the selection criteria are very loose because they aim to maximize the number of member stars and hence only employ a minimum control of contamination.

\citetalias{Li2020d} reported that the actual dispersion of OCs' main sequence is broader than the Gaia formal error.
Taking this into account, we add an additional 0.01 mag to the observational errors of each star in the fitting sample.

\subsection{Field model}
\label{sec:res:fs_model} 
Assuming the photometric distribution of field stars in the sky area is independent of kinematics, we further select a separate field-star sample that is located in the same sky region and within the same magnitude and parallax ranges of the fitting sample but differs from the cluster members in the proper-motion space by $\Delta \mu > 6 \sigma_{\mu, \mathrm{cl}}$. For each cluster, we then construct a nonparametric field model in the CMD based on its field sample following the procedure in Section \ref{sec:field model}.

Taking NGC 2447 as an example, we demonstrate the procedure for the sample selection in Figure \ref{fig:oc_sample}. Panel (a) shows the proper motion of the stars within $r_\mathrm{max}=3\,r_{50}$ from the cluster center. The stars with $\Delta \mu < 4 \sigma_{\mu, \mathrm{cl}}$ (solid circle) are selected as the fitting sample, and the stars with $\Delta \mu > 6 \sigma_{\mu, \mathrm{cl}}$ (dashed circle) are used to build the field model. Their distribution in the CMD is shown in panel (b). The fitting sample contains substantial field stars that deviate from the main sequence, especially in the lower-left (i.e., fainter bluer) region, and these field stars can be well described by the field model (the underlying gray shade), as expected.

\subsection{Prior information}
\label{sec:feh_prior}

Taking reasonable prior can break the parameter degeneracy. Here we fix the power-law index of the binary mass-ratio distribution to the typical value $\gamma_q=0$ (e.g., \citetalias{Li2020d} and references therein). Considering that the spectroscopy [Fe/H] measurement is more reliable than the photometric one, we take the [Fe/H] obtained from previous spectroscopy measurement as a prior \citep{Netopil2016a,Carrera2019a,Donor2020,2021MNRAS.503.3279S}. Specifically, we use a Gaussian distribution $\mathcal{N}(\mathrm{[Fe/H]_P}$, $\sigma_{\mathrm{[Fe/H]_P}})$ truncated to the interval $[-2.3, 0.5]$. When there are multiple [Fe/H] measurements in the literature for one cluster, we use their weighted mean $\mathrm{[Fe/H]_P}$ and corresponding uncertainty $\sigma_{\mathrm{[Fe/H]_P}}$ \citep[following][]{2021npaa.book.....S}. Their values are provided in the last two columns of Table \ref{tab:oc_res}. For the rest of the parameters, we use flat priors within the corresponding ranges that are summarized in Table \ref{tab:model-paras}.

\subsection{Results}
\label{sec:Results}

%----
 \begin{figure*}[!htbp]
  \centering
  \includegraphics[width=1\textwidth]{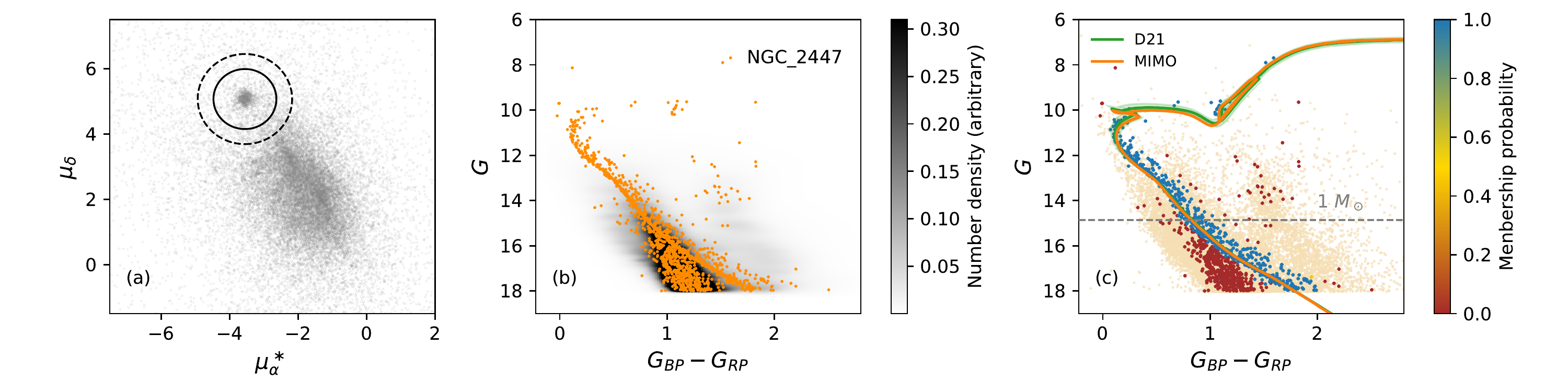}
  \caption{Data selection and fitting result for NGC 2447. 
  (a) Proper motion of the stars within $3r_{50}$ from the cluster center. The stars with $\Delta \mu < 4 \sigma_{\mu, \mathrm{cl}}$ (solid circle) are selected as the fitting sample, and the stars with $\Delta \mu > 6 \sigma_{\mu, \mathrm{cl}}$ (dashed circle) are used to build the field model, where $\Delta \mu$ is the distance to NGC 2447's average proper motion and $\sigma_{\mu, \mathrm{cl}}$ is the dispersion of the cluster (see Section \ref{sec:data_select}). (b) Fitting sample in the CMD. The stars in the fitting sample are shown as orange dots, while the field model is shown as the underlying gray shades. (c) Best-fit result of NGC 2447 in CMD. The fitting sample is colored by the photometric membership probability predicted by MiMO: The high-membership stars (blue) match well with our best-fit isochrone and its uncertainty band (orange curve with shadow), while the low-membership stars, more likely to be field stars (red), follow the same distribution as the field sample (underlying pale yellow dots) as expected. The green curve with shadow shows the isochrone and its uncertainty band fitted by \citetalias{Dias2021a} for comparison. 
  }
  \label{fig:oc_sample}
\end{figure*}
%----

%----
 \begin{figure*}[!htbp]
  \centering
  \includegraphics[width=1.0\textwidth]{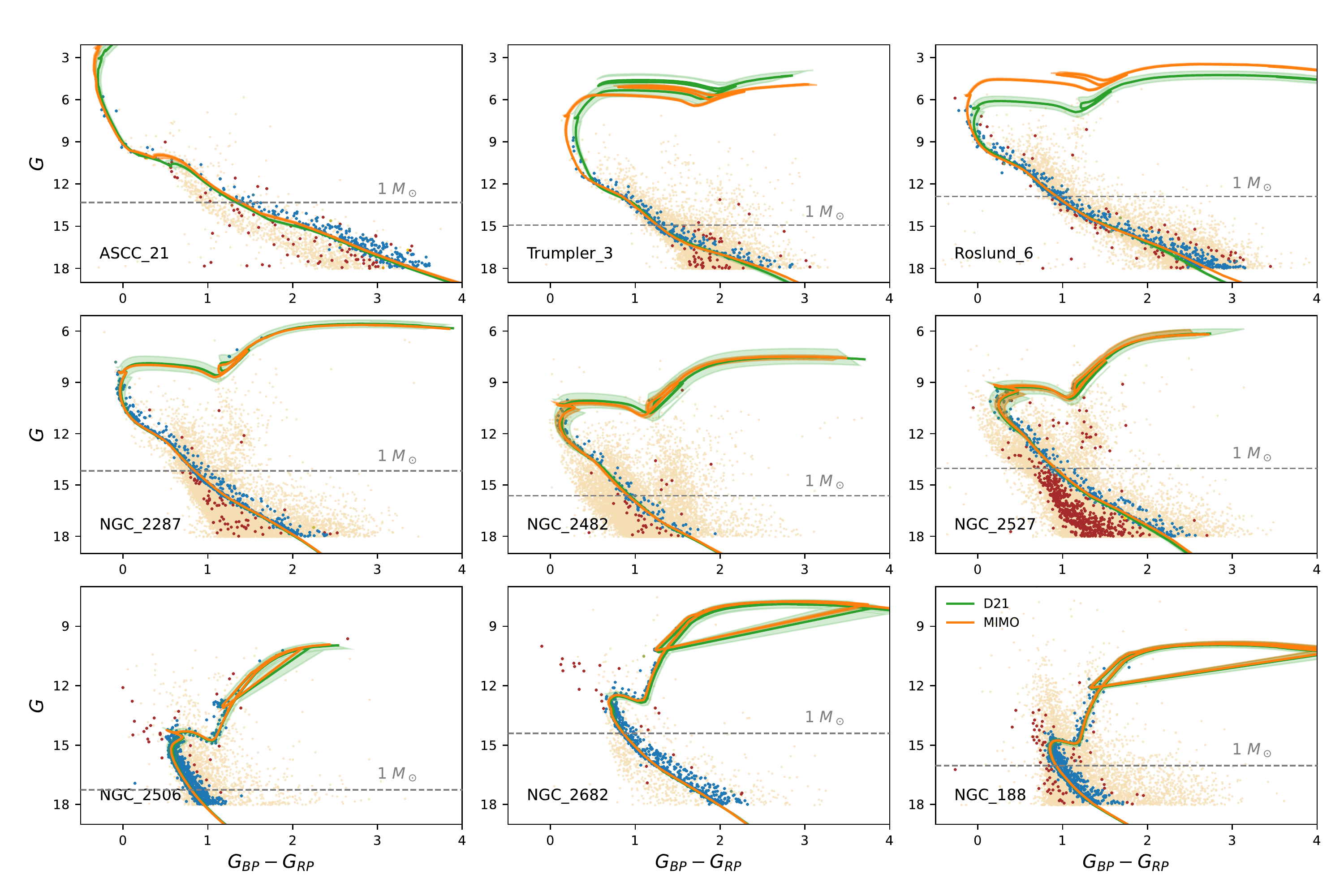}
  \caption{Best-fit result in the CMD, same as the panel (c) of Figure \ref{fig:oc_sample} but for the other nine OCs (sorted by age in ascending order).
  }
  \label{fig:res_cmd}
\end{figure*}
%----

\begin{table*}[bt!]
\caption{Fitting Results of sample clusters}
\label{tab:oc_res}
\footnotesize
\begin{center}
\setlength{\tabcolsep}{3.5pt}
\begin{tabular}{lrrrrrrrrrrrrrrr|rr}
\toprule
\toprule
   OC & logAge & $\sigma_\mathrm{logAge}$ & DM & $\sigma_\mathrm{DM}$ & $A_V$ & $\sigma_{A_{V}}$ &  $\alpha_\mathrm{MF}$ & $\sigma_{\alpha_\mathrm{MF}}$ &  $f_b$ & $\sigma_{f_b}$ & [Fe/H] & $\sigma_\mathrm{[Fe/H]}$  & $f_\mathrm{fs}$ & $\sigma_{f_\mathrm{fs}}$ & $N_\mathrm{tot}$
   &  $\mathrm{[Fe/H]}_P$ & $\sigma_{\mathrm{[Fe/H]}_P}$ \\
% units
   & [dex]  & [dex]  & [mag] &   [mag] &  [mag] &   [mag] & & &  &  & [dex]  & [dex] & & & & [dex]  & [dex] \\
\midrule     
ASCC~21 & 6.982 & 0.025 & 7.542 & 0.042 & 0.197 & 0.017 & $-1.94$ & 0.06 & 0.40 & 0.03 & $-0.157$ & 0.005& 0.28& 0.02& 481& $-0.18$ & 0.01\\
Trumpler~3 & 7.729 & 0.011 & 9.110 & 0.015 & 0.921 & 0.016 & $-2.74$ & 0.14 & 0.46 & 0.05 & 0.141 & 0.007& 0.32& 0.03& 381& 0.13 & 0.01\\
 Roslund~6 & 7.977 & 0.011 & 7.756 & 0.015 & 0.182 & 0.010 & $-1.84$ & 0.07 & 0.41 & 0.03 & 0.048 & 0.007& 0.38& 0.03& 774& 0.01 & 0.02\\
  NGC~2287 & 8.526 & 0.008 & 9.214 & 0.006 & 0.110 & 0.008 & $-2.09$ & 0.08 & 0.30 & 0.03 & $-0.077$ & 0.004& 0.15& 0.01& 764& $-0.11$ & 0.01\\
  NGC~2482 & 8.761 & 0.022 & 10.545 & 0.035 & 0.127 & 0.036 & $-1.96$ & 0.20 & 0.44 & 0.06 & 0.062 & 0.057& 0.26& 0.03& 228& $-0.07$ & 0.10\\
  NGC~2447 & 8.875 & 0.005 & 9.985 & 0.005 & 0.068 & 0.006 & $-2.16$ & 0.10 & 0.35 & 0.03 & $-0.074$ & 0.004& 0.46& 0.02& 1411& $-0.07$ & 0.01\\
  NGC~2527 & 8.972 & 0.026 & 8.861 & 0.038 & 0.142 & 0.020 & $-1.61$ & 0.14 & 0.35 & 0.04 & $-0.040$ & 0.053& 0.66& 0.02& 993& $-0.10$ & 0.04\\
  NGC~2506 & 9.308 & 0.002 & 12.551 & 0.006 & 0.281 & 0.003 & $-2.05$ & 0.16 & 0.43 & 0.03 & $-0.287$ & 0.004& 0.06& 0.01& 1522& $-0.23$ & 0.05\\
  NGC~2682 & 9.590 & 0.003 & 9.632 & 0.006 & 0.093 & 0.011 & $-0.86$ & 0.13 & 0.49 & 0.03 & 0.040 & 0.011& 0.05& 0.01& 970& 0.01 & 0.02\\
   NGC~188 & 9.803 & 0.002 & 11.306 & 0.006 & 0.239 & 0.017 & 0.25 & 0.26 & 0.48 & 0.03 & 0.083 & 0.023& 0.15& 0.01& 1049& 0.10 & 0.01\\
\bottomrule
\bottomrule
\end{tabular}
\end{center}
\tablecomments{--- The first 14 data columns are fitted parameters (described in Table \ref{tab:model-paras}) with their uncertainties. 
$N_\mathrm{tot}$ is the number of stars in the fitting sample, listed for reference. $\mathrm{[Fe/H]}_P$ and $\sigma_{\mathrm{[Fe/H]}_P}$ represent the mean and uncertainty of the [Fe/H] prior (see Section \ref{sec:feh_prior}).}
\end{table*}
%----

We apply MiMO to the 10 OCs. The inferred cluster parameters and their uncertainties for each cluster are summarized in Table \ref{tab:oc_res}. 

We also derive the membership probability for each star, $p_\mathrm{memb}$ (see Equation \ref{eq:pmem}). 
We show the fitting sample of NGC 2447 in the CMD colored by $p_\mathrm{memb}$ in Figure \ref{fig:oc_sample} (c) as an example. Clearly, high-$p_\mathrm{memb}$ stars make up the main sequence and match well with the best-fit isochrone, while low-$p_\mathrm{memb}$ stars follow a distribution similar to that of the field sample. We further confirm that most of the kinematic members selected by \citetalias{Dias2021a} indeed have high photometric membership probability (see Section \ref{sec:phot_member}).

The consistency between the best-fit model and data can also be seen in Figure \ref{fig:res_cmd} for the other nine OCs, which have diverse ages, distances, and field-star contamination levels. This demonstrates the general ability and robustness of our method.

We find that the MF slope is $-2.7$ to $-1.6$ for clusters younger than 2 Gyr, while older clusters appear to have significantly flatter MFs. The binary fraction of these 10 OCs is about 30\% to 50\%. \citetalias{Li2020d} reported a binary fraction of $27\pm 2\%$ in NGC 3532, which is also aligned with this interval. The values are also consistent with those of field stars in the Milky Way \citep{Liu2019a}.

Interestingly, we see blue straggler stars in the three oldest clusters, including NGC~2506, NGC~2682, and NGC~188. These stars are brighter and bluer than the main-sequence turnoff and are also distinguished from the distribution of field stars. These stars have low likelihood of belonging to the mixture model but high kinematic membership probabilities. This is because our cluster model does not cover the blue straggler component. It is possible to extend MiMO with an additional blue straggler stars model in the CMD to find and characterize them.

%==============================================================
\section{Discussion}
\label{sec:disscuss}
In this section, we compare our results with the literature and discuss the advantages, limitations, and possible extensions of our method.

\subsection{Comparison with previous work}
\label{sec:compare_d21}

%----
 \begin{figure*}[!htbp]
  \centering
  \includegraphics[width=1\textwidth]{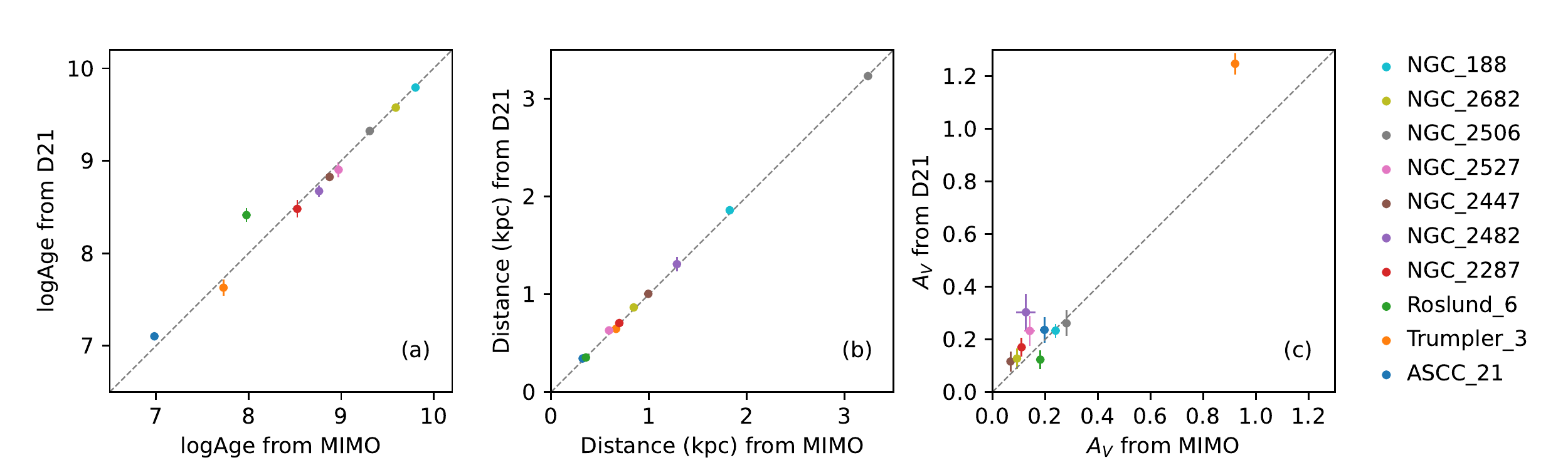}
  \caption{Comparison of (a) logAge, (b) distance, and (c) $A_V$ measured by MiMO and \citetalias{Dias2021a}. Each color represents an OC as indicated in the legend. Error bars show the fitting uncertainties provided by each method, some of which are smaller than the symbol size.}
  \label{fig:compare_d21}
\end{figure*}
%----

%----
 \begin{figure*}[!htbp]
  \centering
  \includegraphics[width=1\textwidth]{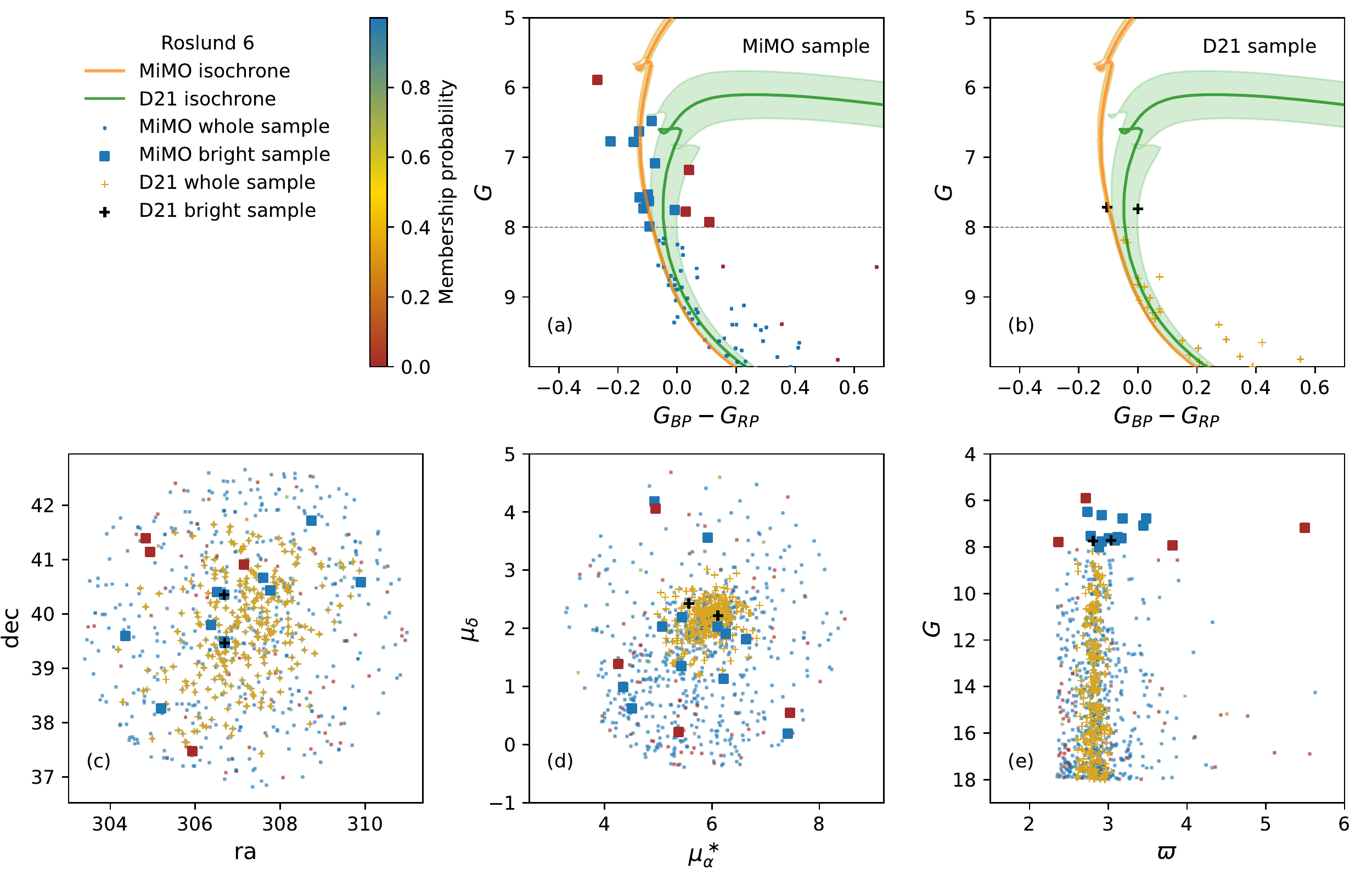}
  \caption{Comparison of the fitting sample used by MiMO and that by \citetalias{Dias2021a} for Roslund 6. Panels (a) and (b): 
  the fitting sample and best-fit isochrone in the CMD used by MiMO and \citetalias{Dias2021a}, respectively. Panels (c), (d), and (e): fitting sample in the space of sky coordinates, proper motion, and parallax vs. magnitude. Our sample is shown by squares color-coded by MiMO membership probability (see color bar), and the \citetalias{Dias2021a} sample is shown by yellow crosses. The bright stars with $G < 8$ mag in each sample are further highlighted by different symbols as indicated in the legend.}
  \label{fig:roslund6}
\end{figure*}
%----

We compare our best-fit cluster parameters, including age, distance, and dust extinction, with those measured by \citetalias{Dias2021a} in Figure \ref{fig:compare_d21}. The corresponding isochrones from these two methods are shown in Figure \ref{fig:res_cmd}. \citetalias{Dias2021a} inferred the OC properties using Gaia DR2 photometry based on a sample of high-membership probability. Overall, we find a consistency between our results and those of \citetalias{Dias2021a}, especially in the photometric distance estimate, while our reported uncertainties are much smaller. The moderate difference in dust extinction also seems reasonable compared with the measurement uncertainty.

It is worth looking into the clusters that show a large discrepancy in parameter estimates. For example, our result suggests a significantly younger age for Roslund 6 than that by \citetalias{Dias2021a} (see Figure \ref{fig:compare_d21}). Comparing panels (a) and (b) of Figure \ref{fig:roslund6}, we find that our fitting sample contains much more stars in the main-sequence turnoff region than that of \citetalias{Dias2021a}. Specifically, there are 11 stars with a high membership probability derived by MiMO that are brighter than $G=8$ mag in our sample, while there are only 2 in the \citetalias{Dias2021a} sample. According to the distribution of space, proper motion, and parallax shown in panels (c), (d), and (e), the majority of these 11 stars are indeed likely to be cluster members. However, \citetalias{Dias2021a} has discarded most of them due to a more stringent sample selection, leading to an overestimate of the age of Roslund 6.

A similar discrepancy can be seen when comparing the isochrones of NGC 2447, Trumpler 3, and NGC 2682 in Figure \ref{fig:oc_sample} and \ref{fig:res_cmd}, where our best-fit isochrones clearly match with the data better than those of \citetalias{Dias2021a}.

As demonstrated above, enabling a natural treatment of field contamination, and hence allowing a sample containing more complete cluster members including more bright stars at the main-sequence turnoff region, is a particular advantage of our mixture model. The determination of cluster age is very sensitive to such bright stars. Therefore, our method enables a better parameter determination in terms of both precision and robustness.

\subsection{Photometric membership probability}
\label{sec:phot_member}

 \begin{figure}[!htbp]
  \centering
  \includegraphics[width=0.5\textwidth]{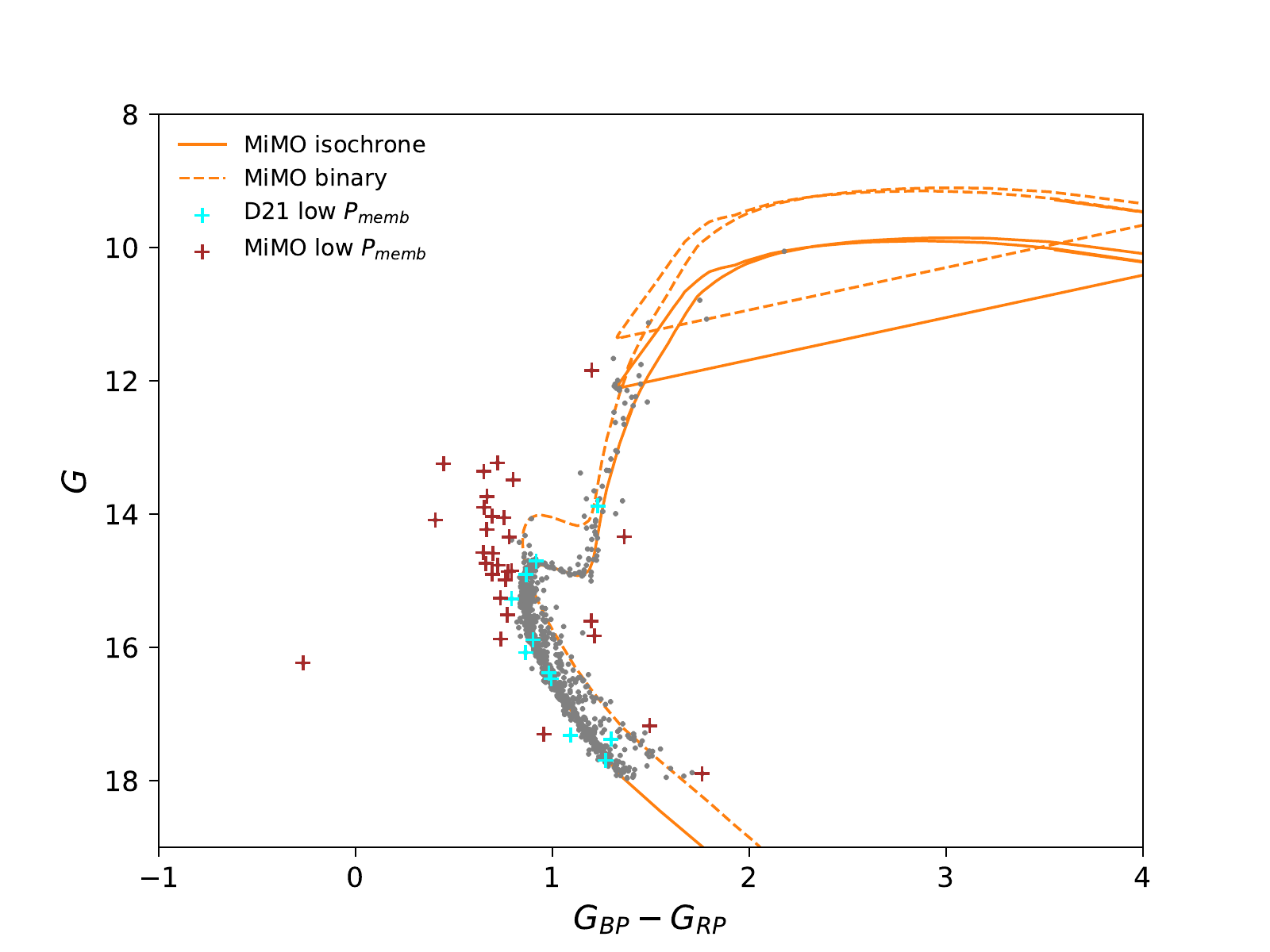}
  \caption{Comparison of membership probability obtained from MiMO and \citetalias{Dias2021a} for NGC 188. Gray dots represent the stars in common in the two samples in the CMD. Stars with $p_\mathrm{memb, D21}>0.5$ but $p_\mathrm{memb, MiMO}<0.5$ are shown as brown crosses, while stars with $p_\mathrm{memb, D21}<0.5$ but $p_\mathrm{memb, MiMO}>0.5$ are shown as blue crosses.}
  \label{fig:p_compare}
\end{figure}

We compare the membership determination of MiMO with that of \citetalias{Dias2021a} (Gaia DR2 kinematics based) and find good consistency between photometric and kinematic $p_\mathrm{memb}$ in general.

Taking NGC 188 as an example, the \citetalias{Dias2021a} sample contains 883 stars with kinematic $p_\mathrm{memb}$ > 0.2, 875 of which are also presented in our sample. Among the stars in common, 784 stars have $p_\mathrm{memb}>0.8$ and 830 stars have $p_\mathrm{memb}>0.5$ in both methods.

There are 30 stars with $p_\mathrm{memb, D21}>0.5$ but $p_\mathrm{memb, MiMO}<0.5$. As shown by the brown crosses in Figure \ref{fig:p_compare}, these stars are located away from the main sequence in the CMD, which suggests that they are possibly either field stars with similar kinematics by coincidence or the blue straggler members.
Conversely, there are 11 stars with $p_\mathrm{memb, MiMO}>0.5$ but $p_\mathrm{memb, D21}<0.5$. As shown by blue crosses in Figure \ref{fig:p_compare}, they are well located on the main sequence or the binary sequence, which explains their high photometric membership probability in MiMO.

Our method provides an independent determination of membership probability. As one important application, it enables us for search candidates of blue stragglers (or other special stars not covered by the simple stellar population model) by looking for stars with high kinematic $p_\mathrm{memb}$ but low photometric $p_\mathrm{memb}$.

\subsection{Photometric distance}
\label{sec:photd}
%----
 \begin{figure}[!htbp]
  \centering
  \includegraphics[width=0.3\textwidth]{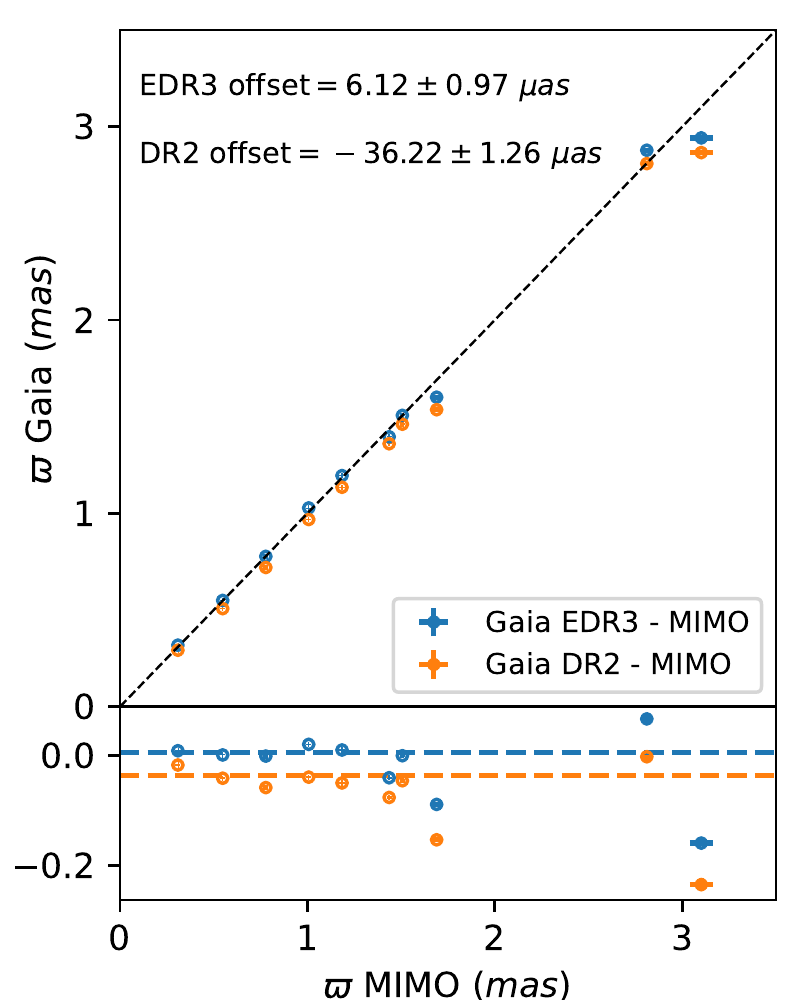}
  \caption{Comparison between the OC parallaxes inferred from this work (MiMO, photometric distance) and those from Gaia astrometry (DR2 and EDR3). The differences and average offsets between Gaia's and our values are shown in the lower panel as symbols and dashed lines, respectively.}
  \label{fig:phot_plx}
\end{figure}
%----

We show the comparison between the parallax derived from our photometric distance and that directly obtained from Gaia EDR3 (ZY Shao, 2022, in preparation). \citet{Shao2019} established a mixture model to derive the parallax of the globular clusters with Gaia DR2. They applies the same method to the OCs based on Gaia EDR3 with the zero-point correction of individual stars \citep{Lindegren2021}.

As demonstrated in Figure \ref{fig:phot_plx}, we see a notable agreement between MiMO and Shao (2022, in preparation). The consistency between these two independent methods based on different data spaces suggests that they are both reliable. On the other hand, Gaia EDR3's parallax zero-point correction is reasonable.

\citet{Cantat-Gaudin2020b} presented the mean parallax and its dispersion of OCs based on their members. We also show the comparison between our photometric-distance-derived parallaxes and parallaxes for OCs in using Gaia DR2 \citep{Cantat-Gaudin2020b} but without a zero-point correction. We can see an offset of $-36.22\ \mu as$ between the two (While this is only $6.12\ \mu as$ with EDR3.). This offset is marginally agreed with the mean parallax zero-point value in Gaia DR2 \citep{Arenou2018a,Shao2019}.

The differences between the parallax derived from DR2 and EDR3 are caused by two reasons. One is that the accuracy of EDR3 is higher than that of DR2, and the other is that the parallaxes obtained from DR2 are not zero-point corrected.

It is worth pointing out that there is a dispersion of the parallax zero point of about $13\sim15\ \mathrm{\mu as}$ \citep{Lindegren2021} in Gaia EDR3. For the distant OCs, they have much smaller parallaxes, and this zero-point dispersion can be significant. In contrast, the uncertainty in the photometric distance mostly relies on the sample size (number of stars). Therefore, the accuracy is consistent regardless of the distance. Thus, for distant OCs, the photometric distance is still an very important measurements, and can be more accurate than the distance obtained from the parallax.

\subsection{Validation of the mass function}
\label{sec:mf}

 \begin{figure*}[!htbp]
  \centering
  \includegraphics[width=1.0\textwidth]{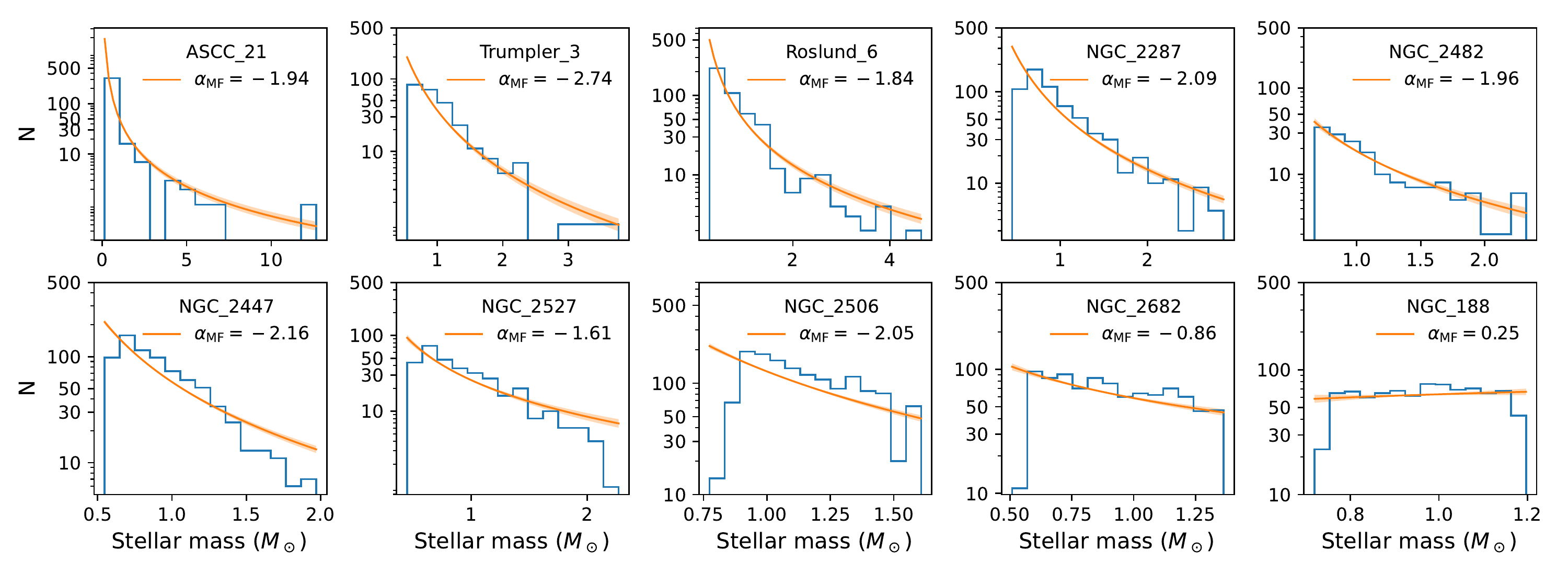}
  \caption{Sanity check of the best-fit MF. In each panel, the orange line with the shadow shows the MF and its associated uncertainty determined by MiMO. For comparison, the histogram shows the distribution of stellar mass directly obtained from data.
  }
  \label{fig:all_mf}
\end{figure*}
%----

For validation of the assumed power-law MF and the corresponding inferred slope, we compare our best-fit MF with the MF directly counted from the data in Figure \ref{fig:all_mf}. We sort stars in the fitting sample by their membership probability from high to low and label the first $N_\mathrm{cl}=(1-f_\mathrm{fs})N_\mathrm{tot} $ stars as cluster members. For each member star, we then estimate its mass (or the mass of the major component if it is a binary) by searching the closest point on the single and binary sequences. The resulting mass distributions for stars above $G=18$ mag in the 10 clusters are shown as histograms in Figure \ref{fig:all_mf}.
In all cases, the MF data match well with the best-fit MF from MiMO.

Note that the above calculation of the direct MF is approximate and mainly for illustration purposes. Strictly speaking, the membership should be treated in a probabilistic manner rather than as a simple selection, and the observational errors should be considered as well. As a completely Bayesian method, MiMO can naturally predict the mass of member stars (and the mass ratio for binaries) through the hierarchical Bayesian technique in a rigorous and self-consistent way. We leave such exploration to future work.

\subsection{Limitation and possible extension}
It should be mentioned that the measurement of the binary fraction and the MF slope suffers from the slight mismatch between theoretical stellar models and observed main sequences (as discussed in \citetalias{Li2020d}). Their best-fit values provided by MiMO only represent the average performance of the overall distribution in the CMD. More detailed studies on binary properties and mass function can benefit from a precise empirical measurement of the main-sequence ridge line and its scatter as done in \citetalias{Li2020d} and \citealt{Li2021h}.

As a framework, MiMO has great potential to extend its ability. It is easy to incorporate other stellar models, astrophysical processes, and alternative functional forms of the MF and binary ratio function (e.g. log-normal or tapped power law). For instance, we can use a more complicated stellar evolution model with the stellar rotation included to constrain the stellar rotation distribution in a stellar clusters. We can also model the blue straggler stars or multiple stellar populations as extra components in the mixture distribution in the CMD. Because MiMO is designed under the Bayesian framework, it is also convenient to assess the different models and choose the number of mixture components automatically.

\section{Conclusion}
\label{sec:Conclusion}

In this work, we have developed a state-of-the-art Bayesian framework named MiMO to measure the OC parameters from photometric data. MiMO treats an open cluster in the CMD as a mixture of single stars, unresolved binaries, and contaminated field stars. The model of the first two components is built based on the theoretical stellar evolution model, stellar MF and binary properties, while the model of field stars is constructed nonparametrically from a separate field-star sample in the vicinity of the cluster. The comprehensive physical models and advanced statistics enable us to measure the OC properties, including isochrone parameters (age, distance, metallicity, and extinction), stellar MF, and binary parameters (binary fraction and mass-ratio distribution) precisely at the same time.

Based on the Gaia EDR3 photometry, we demonstrate the high internal precision of MiMO with mock clusters. For a mock cluster of $\sim\! 1000$ member stars with less than 50\% field contamination, MiMO can infer the logAge, distance modulus, and dust extinction within 0.02 dex, 0.02 mag, and 0.02 mag, respectively, which presents an improvement of a factor of 10 compared with previous work.

We then apply MiMO to 10 real OCs with Gaia EDR3 data. Their best-fit parameters are listed in Table \ref{tab:oc_res}. The MF slope is $-2.7$ to $-1.6$ for clusters younger than 2 Gyr, while older clusters appear to have significantly flatter MFs. The binary fraction is 30\% to 50\%, consistent with that of field stars in the Milky Way \citep{Liu2019a}.

The photometric distances are fairly consistent with the astrometric distance. The best-fit age, distance, and extinction are in good agreement with previous measurements by \citetalias{Dias2021a} in general. However, \citetalias{Dias2021a} appears to overestimate the age or metallicity for several OCs. It is because the stringent membership selection of \citetalias{Dias2021a} has removed most of the bright stars around the turnoff region that are crucial for age estimation. In contrast, our method does not rely on ``pure'' members but treats the field contamination as a model component, which hence allows a sample containing more complete cluster members. This not only has the statistical advantage of the sample size but also minimizes the bias in age by retaining more bright stars. Therefore, our method presents a better parameter determination in terms of both precision and robustness.

Besides the precision and robustness, MiMO has several main advantages:

1.~It does not rely on a strict membership determination and is hence applicable to pure photometric data as well.

2.~As a Bayesian framework, it can naturally incorporate other independent measurements as priors, e.g., metallicity and extinction from spectroscopic data and distance from astrometry data.

3.~It is straightforward to extend with other models of physical processes or stellar components, e.g., the broadening of main sequence due to stellar rotation or the additional distribution of blue straggler stars in the CMD, when available.

4.~It can output the photometric membership probability, $p_\mathrm{memb}$. As one important application, the stars with high kinematic $p_\mathrm{memb}$ but low photometric $p_\mathrm{memb}$ represent candidates of blue stragglers and other special stars that are not covered by the single-stellar-population model.

In the Gaia era, our method can maximize the exploitation of precise photometric data. Combining MiMO and Gaia data, we plan to measure the parameters of all Galactic OCs in the future. Such a large catalog with high precision will be particularly useful for understanding the formation and evolution of OCs and for tracing the structure and history of the Milky Way disk.

% \ 

%==============================================================
\begin{acknowledgments}
We sincerely thank the anonymous referee for valuable comments and suggestions. We thank Zhaozhou Li for the careful proofreading and Li Chen, Jing Zhong, Chao Liu, and Wenping Chen for the helpful discussions. This work is supported by National Key R\&D Program of China No.\ 2019YFA0405501, the National Natural Science Foundation of China (NSFC) under grant U2031139, and the science research grants from the China Manned Space Project with No. CMS-CSST-2021-A08. L.L. thanks Avishai Dekel for hosting her visit at the Hebrew University of Jerusalem under the support of the UCAS Joint PHD Training Program. This work has made use of data from the European Space Agency (ESA) mission Gaia (\url{https://www.cosmos.esa.int/gaia}), processed by the Gaia Data Processing and Analysis Consortium (DPAC; \url{https://www.cosmos.esa.int/web/gaia/dpac/consortium}). Funding for the DPAC has been provided by national institutions, in particular the institutions participating in the Gaia Multilateral Agreement.
%==============================================================

\software{Astropy \citep{AstropyCollaboration2013}, PARSEC \citep{Bressan2012b}, Numpy \citep{vanderWalt2011}, Scipy \citep{Oliphant2007}, Matplotlib \citep{Hunter2007}, {dynesty} \citep{Speagle2020}.}
\end{acknowledgments}

\appendix

\section{Numerical implementation}
\label{sec:implementation}

While the general methodology is clear, implementing it with satisfying precision and reasonable CPU/memory usage is an intractable task. The likelihood function depends on each parameter (except for $f_\mathrm{fs}$) in a complicated way, e.g., through multiple integrals involving the mass function, binary mass-ratio distribution, and observation errors. The keys to the solution are to (1) use weighted summation of discrete points instead of integrals and (2) find a balance between the use of the precomputed library (memory) and runtime computation (CPU time).

We plan to make our code public in the near future.

\subsection{Isochrone model library}
\label{sec:model_lib}

We construct a theoretical isochrone library for parameters on a 3D grid of (logAge, $\mathrm{[Fe/H]}$, $A_V$). Each dimension of the grid is required to cover the parameter range listed in Table \ref{tab:model-paras}. The grids of logAge and $\mathrm{[Fe/H]}$ are equally spaced by 0.05 dex and 0.05 dex, respectively, while the grid of $A_V$ is taken to be $[0, 0.5, 1, 2, 5, 10]$. The same interpolating grid of $A_V$ has been adopted by the original YBC extinction model \citep{Chen2019c}; therefore, using a denser $A_V$ grid here will not increase the precision but will increase the memory cost. For each parameter set in the library, we download the isochrone from the web interface of the PARSEC stellar model (with the YBC extinction model and Gaia EDR3 photometric system integrated) with a Python script\footnote{Written by Zhaozhou Li, first used in \citetalias{Li2020d}, and publicly available at \url{https://github.com/syrte/query_isochrone}.}. Each raw isochrone consists of the $G$, $\Gbp$, and $\Grp$ absolute magnitudes (and hence $\BPRP$) for a table of initial stellar mass $\mathcal{M}$. We denote the CMD of $(G, \BPRP)$ by $(M, c)$ hereafter. 

The raw isochrones are not dense enough for a discrete integral. For an isochrone with $A_V=0$, we refine the $\mathcal{M}$ table by inserting additional points and interpolating $(M,c)$ over $\mathcal{M}$ linearly, so that the separation in $M$ or $c$ between any two adjacent points is smaller than 0.005 mag (about half of the main-sequence scatter, see \citetalias{Li2020d}). For each interval in the refined table, we then record the mass range $[\mathcal{M}_i^-, \mathcal{M}_i^+]$, the midpoint mass $\mathcal{M}_i$, and the absolute magnitude and color $(M_i, c_i)$ at $\mathcal{M}_i$. Given logAge and $\mathrm{[Fe/H]}$, for an isochrone with $A_V>0$, we interpolate over the same refined mass table as the case $A_V=0$. In this way, we turn each isochrone curve (for single stars) into a set of model points with $N\sim 5000$ respectively.

The next step is to prepare the model for binaries. For each isochrone in the library, we generate eight associated binary sequences with binary ratios, $q=[0.25, 0.35, \ldots, 0.95]$. Note that each sequence actually represents a $q$ interval, e.g., $q=0.25$ represents $[q^-, q^+]=[0.2, 0.3]$. Following the same procedure with the same refined mass table as single stars, we turn each binary sequence into a set of model points as well. The $(M_i, c_i)$ pair of a binary point is obtained by combining two stars with $\mathcal{M}_i$ and $q\mathcal{M}_i$.

As a summary of the above, each isochrone in the library grid is converted into and stored in the disk as a set of discrete (but dense) model points, $\{M_i, c_i$, $\mathcal{M}_i^-, \mathcal{M}_i^+$, $[q_i^-, q_i^+]\}_{i=1,\ldots,N}$ with $N\sim 5000\times 9$, including a single sequence ($q=0$) and eight binary sequences ($q \in [0.2, 1]$). Considering the additional smoothing due to observation error, the above discretization seems to be a good compromise between precision and computing efficiency. 

\subsection{Member-star distribution}

Providing the simple stellar population parameters ($\Theta_\mathrm{cl}$): logAge, $\mathrm{[Fe/H]}$, $A_V$, distance module (DM), mass function ($\mathcal{F}_{\mathrm{MF}}$), binary ratio ($\fb$), and binary mass function ($\mathcal{F}_q$), we calculate the likelihood of being a member star. Here we consider three cases separately, depending on whether logAge, $\mathrm{[Fe/H]}$, and $A_V$ are on grid nodes of the stored model library.

\textit{Case 1}: logAge, $\mathrm{[Fe/H]}$, and $A_V$ all on the grid node.
On the basis of Appendix \ref{sec:model_lib}, 
we represent the continuous model distribution $\phi_\mathrm{cl}(m, c)$ (Equation \ref{eq:phi-cl})
by a set of discrete model points $\{m_i=M_i+\mathrm{DM}, c_i\}$ with weight $w_i$, where
\begin{equation}
    w_i = (1 - \fb)  \int_{\mathcal{M}_i^-}^{\mathcal{M}_i^+}
    \mathcal{F}_{\mathrm{MF}} (\mathcal{M}) d\mathcal{M}
\label{eq:wi1}
\end{equation}
for a single model point or
\begin{equation}
    w_i = \fb \int_{\mathcal{M}_i^-}^{\mathcal{M}_i^+} \mathcal{F}_{\mathrm{MF}}
    (\mathcal{M}) d\mathcal{M} \times \int_{q_i^-}^{q_i^+} \mathcal{F}_q (q) dq
\label{eq:wi2}
\end{equation}
for a binary model point. 

For an observed star $(m_\mathrm{ob}, c_\mathrm{ob})$ with observation error $(\sigma_{m},\sigma_{c})$, the likelihood of being a member star (Equation \ref{eq:obs_psi}) becomes
\begin{equation}
    \psi_\mathrm{cl}(m_\mathrm{ob},c_\mathrm{ob}) = \mathcal{C} \sum_i w_i 
       \mathcal{N}(m_\mathrm{ob} \,|\,M_i + \mathrm{DM},\sigma_{m}) \mathcal{N}(c_\mathrm{ob}\,|\,c_i,\sigma_{c}),
\label{eq:dis_psi}
\end{equation}
where the normalization factor $C$ is computed within the observation magnitude limit $[m_1, m_2]$,
\begin{equation}
    C = \sum_i w_i \int_{m_1}^{m_2} \mathcal{N}(m \,|\, M_i+\mathrm{DM},\sigma_{m}) d m.
\label{eq:app_norm}
\end{equation}

\textit{Case 2}: Only logAge and $\mathrm{[Fe/H]}$ on the grid node. We first interpolate the model points $(m_i, c_i)$ linearly for $A_V$ from the two adjacent $A_V$ nodes (which share the same refined mass table) with the same age and metallicity. The rest is the same as Case 1.

\textit{Case 3}: The general case in practice. The likelihood of an observed star is bilinearly interpolated according to the four adjacent nodes in (logAge, $\mathrm{[Fe/H]}$), whose likelihoods are computed as Case 2. We thus avoid the complexity of direct interpolation between isochrones.

\subsection{Field-star distribution}
\label{sec:field_kde}

We build the model of field-star distribution in the CMD from a sample of neighboring field stars, $\{m_j, c_j\}_{j=1,...,N_\mathrm{fs}}$, through the kernel density estimation. Combining Equation (\ref{eq:field stars_prob}) and (\ref{eq:obs_psi}), the likelihood of being a field star for observed $(m_\mathrm{ob}, c_\mathrm{ob})$ is
\begin{align}
    \psi_\mathrm{fs}(m_\mathrm{ob}, c_\mathrm{ob}) = \frac{C}{N_\mathrm{fs}} \sum_{j=1}^{N_\mathrm{fs}} \mathcal{N}(m_\mathrm{ob} \,|\, m_j,\epsilon_{m,j})\mathcal{N}(c_\mathrm{ob} \,|\, c_j,\epsilon_{c,j}),
\end{align}
where the normalization factor $C$ is computed similarly to Equation (\ref{eq:app_norm}) and the smoothing sizes, $\epsilon_{m,j}$ and $\epsilon_{c,j}$, are determined as follows (see also \citealt{Li2019d}). For each star (labeled as $j$) in the field sample, we calculate the distance, $\epsilon_{j}$, to its $k$th nearest neighbor in the CMD (with $k=\sqrt{N_\mathrm{fs}}$). Then we adopt $\epsilon_{m,j}=({s\epsilon_{j}^2+\sigma_{m,j}^2})^{1/2}$ and similarly for $\epsilon_{c,j}$, where $s=1$ is a factor to control the smoothness and $\sigma_{m,j}$ is the observational error of this star (recall that the fitting sample and field sample share the same observation error distribution). In this way, the smoothing size of a field star is larger if it resides in a low-density region of the CMD or it has greater observational errors. From our experience, the result is not sensitive to the details of smoothing. Varying the smoothing size by a factor of 2 or scaling the CMD before searching the nearest neighbors leads to a negligible change of the inferred cluster parameters.

\subsection{Likelihood}
\label{sec:app_like}

Given the cluster parameters $\Theta_\mathrm{cl}$ and field-star fraction $f_\mathrm{fs}$,
the mixture distribution of member stars and field stars is
\begin{align}
    \psi(m_\mathrm{ob},c_\mathrm{ob}|\ \Theta) = (1-f_\mathrm{fs}) \psi_\mathrm{cl}(m_\mathrm{ob},c_\mathrm{ob}\,|\ \Theta_\mathrm{cl}) + f_\mathrm{fs}\psi_\mathrm{fs}(m_\mathrm{ob},c_\mathrm{ob})
\end{align}
(see also Equation \ref{eqn:lik_err}), where $\Theta=\{\Theta_\mathrm{cl}, f_\mathrm{fs}\}$.
The likelihood of an observed sample, $\mathcal{D}=\{m_{\mathrm{ob},i},c_{\mathrm{ob},i}\}_{i=1, \ldots, N}$, gives
\begin{equation}\label{eq:lhtot}
    \mathcal{L}(\mathcal{D}|\,\Theta) = \prod_{i=1}^{N}\ \psi(m_{\mathrm{ob},i}, c_{\mathrm{ob},i}|\,\Theta).
\end{equation}

\subsection{Sampling the posterior}
\label{sec:dynesty}

We employ the nested sampling method (\citealt{Skilling2004a,Skilling2006}) implemented by the public package \texttt{dynesty} \citep{Speagle2020} to obtain the posterior distribution of the parameters. For the reader's reference, we use 1000 live points with multiple bounding ellipsoids, sample points through random walks with fixed proposals, and stop the iteration at $d\log z=0.1$ (see the \texttt{dynesty} documentations for their meanings). A typical cluster with 1500 stars in the fitting sample takes $\sim$20 GB of memory (mostly used by the isochrone library) and 5 hr with a single CPU core. Note that the isochrone library can be used for fitting multiple clusters in parallel, so the memory usage does not necessarily increase significantly with the number of clusters.

\bibliographystyle{aasjournal}
\bibliography{main}

\end{CJK*}
\end{document}